\documentclass[preprint]{elsarticle}

\usepackage{amssymb, latexsym, amsthm, amsmath,lineno,epsfig,mathtools,hyperref,todonotes,booktabs,cite}
\usepackage[singlelinecheck=false]{caption}
\modulolinenumbers[1]

\journal{Theoretical Population Biology}

\newcommand\given{{\,|\,}}

\newcommand\eg{{\it e.g.,}}

\newcommand\ie{{\it i.e.,}}

\begin{document}

\begin{frontmatter}
\title{A nearly-neutral biallelic Moran model with biased mutation and linear and quadratic selection}
 
\author[address1,address2]{Claus Vogl\corref{correspondingauthor}}
\cortext[correspondingauthor]{Corresponding author}
\ead{claus.vogl@vetmeduni.ac.at}
\author[address3]{Lynette Caitlin Mikula}
\ead{lcm29@st-andrews.ac.uk}

\address[address1]{Department of Biomedical Sciences, Vetmeduni Vienna, Veterin\"arplatz 1, A-1210 Wien, Austria}
\address[address2]{Vienna Graduate School of Population Genetics, A-1210 Wien, Austria}
\address[address3]{Centre for Biological Diversity, School of Biology, University of St.\ Andrews, St Andrews KY16 9TH, UK}

\begin{abstract}

In this article, a biallelic reversible mutation model with linear and quadratic selection is analyzed. The approach reconnects to one proposed by \citet{Kimu81}, who starts from a diffusion model and derives its equilibrium distribution up to a constant. We use a boundary-mutation Moran model, which approximates a general mutation model for small effective mutation rates, and derive its equilibrium distribution for polymorphic and monomorphic variants in small to moderately sized populations. Using this model, we show that biased mutation rates and linear selection alone can cause patterns of polymorphism within and substitution rates between populations that are usually ascribed to balancing or overdominant selection. We illustrate this using a data set of short introns and fourfold degenerate sites from {\em Drosophila simulans} and {\em Drosophila melanogaster}.
\end{abstract}


\end{frontmatter}

\newpage

\section{Introduction}

Our article draws on a model introduced by \citet{Kimu81}: He considered a normally distributed phenotypic trait that has normal fitness effects and is influenced by multiple biallelic loci. The forces acting on individual sites are expanded up to second order to account for dominance and over- and underdominance. The purpose of Kimura's article is twofold: (i) to show that (nearly-) neutral evolution is possible under stabilizing selection, and (ii) to explain codon usage bias, \ie{} the preferential use of certain codon triplets for identical amino acids or termination signals. His treatment is, however, inconsistent because he switches between models with reversible mutation for modeling polymorphism and irreversible mutation for modeling substitution rates. It is also incomplete because he ignores the influence of biased mutation rates. In this article, we revive the linear and quadratic selection scenario of \citet{Kimu81} and incorporate it into a biallelic boundary-mutation Moran model, which allows for biased, reversible mutations from monomorphic states \citep[][]{Vogl12}. This model is consistent in the sense that it can be used for modeling both polymorphism and substitution rates. With it, we demonstrate the importance of the interplay between directional and quadratic selection and mutation bias. 

According to the neutral theory \citep{Kimu83}, newly mutated alleles are either selectively neutral and thus subject only to random drift, or strongly selected and thus quickly weeded out or fixed. Therefore neutral alleles alone contribute to polymorphism. Nevertheless, Kimura himself proposed and analyzed models that include weak selection. Assuming a single segregating allele that originated by mutation, the probability of fixation can be determined in the presence of linear and quadratic selection \citep[\eg][]{Kimu62}. Later, \citet{Kimu69a} proposed the infinite sites model, in which derived alleles originate by mutation from infinitely many ancestral sites with a given mutation rate. Derived alleles may be favored or disfavored by directional selection. Note that with this model mutations are irreversible. Due to the infinite supply of ancestral alleles, a quasi-equilibrium between mutation, drift and directional selection with constant polymorphism may develop, whereas the fitness increases or decreases indefinitely depending on the direction of selection. Under these assumptions, Kimura derived expressions for site heterozygosity. The Ewens-Watterson estimator of genetic diversity can also be determined with this approach (see Sec.~\ref{section:het}). Later still, \citet{Ohta73} argued for the pervasive occurrence of slightly deleterious mutations to explain the constancy of substitution rates among organisms with different generation times. Her work ultimately led to the nearly-neutral theory \citep{Ohta96}, where the strength of selection and drift is approximately equal. 

The (nearly-)neutral theory sets the stage for comparing polymorphism and substitutions between different classes of mutations with the McDonald-Kreitman test \citep{McDo91}: Mutation of a site within a coding sequence may lead to replacement of an amino acid, and therefore this mutation is likely subject to (weak or strong) selection. However, the mutation may also leave the amino acid unaltered and therefore likely be neutral on this level of selection. Substitution rates are given by the product of mutation rates and fixation probabilities, which depend on the scaled selection strength. Polymorphism is influenced by the same forces, but is less affected by selection than substitutions. Ratios of non-silent (replacement) vs.\ silent substitutions (the $D_n$/$D_s$ ratio) and of non-silent vs.\ silent polymorphism (the $P_n$/$P_s$ ratio) are then reported as part of the testing paradigm. Interestingly, \citet{McDo91} found an excess of replacement substitutions within the alcohol dehydrogenase gene of {\em {Drosophila}}, which seems to indicate positive selection. Later, the $D_n$/$D_s$ ratio was used to infer negative or positive selection of a specific gene or lineage vs.\ neutral \citep{Yang98,YangBial00,Yang02} or neutral and purifying \citep{Zhang05} evolution. In this framework, a debate about the relative proportions of neutral, positive, and negative mutations has developed \citep[\eg][]{Yang00,Smit02,Niel03}. 

Irreversible mutations, as in the infinite sites model, are consistent with the usual model for substitutions \citep{Kimu62}: The trajectory of a single mutation in a population is followed to either fixation or loss with a diffusion approach. Irreversible mutation models do not allow an equilibrium to develop in a population of finite size. Reversible mutation-selection-drift models, however, reach an equilibrium. With a diffusion approach and a general mutation model, the equilibrium distribution for allele frequencies was given by \citet{Wrig31} and often used later \citep[\eg][]{Kimu81,Li87,Bulm91}. In this case, the boundaries are inaccessible for nontrivial mutation parameters, which is not consistent with the above substitution model. With selection, Wright's distribution is defined up to a constant of proportionality that usually needs to be determined numerically.

Assuming small scaled mutation rates, only a single mutation will likely segregate in a small to moderately sized population sample. In this limit, it is possible to derive models that allow for explicit calculation of equilibrium distributions, substitution rates, and other informative quantities by assuming a boundary-mutation Moran model \citep{Vogl12} (see Sec.~\ref{section:historical_intro} and Sec.~\ref{section:model}). We note that the mathematical tractability of the boundary models also enables their use in phylogenetic settings \citep{DeMa13,DeMaio2015}, where recurrent mutations need to be assumed. Using expressions derived from such models, it can be shown that mutation bias and linear (directional) selection of the same magnitude can affect polymorphism and substitutions in surprising ways (see Sec.~\ref{section:statistics}). This interesting interplay has already been demonstrated by \citet{McVe99}, who intuitively combined reversible mutations (with positively and negatively selected alleles) with the infinite sites model. In the appropriate limit, many of their results are identical to those obtained with the boundary-mutation Moran model (as we show throughout our article). Since mutation biases are rarely extreme, the selection strength $\gamma$ acting on them is often within the nearly-neutral range of $0.2< |\gamma| <3$ \citep{Tach91}, where $\gamma=4Ns$ with the diploid Wright-Fisher model and $\gamma=Ns$ with the haploid Moran model. 

Silent mutations seem to be under selective constraint in a wide array of organisms: codon usage bias has been shown to alter the silent substitution rate in mammals and birds \citep{Rou19}, the aspen tree {\em Populus tremula} \citep{Ingv10}, as well as fruitfly species of the genus {\em Drosophila} \citep{Akas94}. \citet{Machado19} and \citet{Lawrie20} have also shown that codon usage bias appears to account for a substantial amount of the total selective pressure acting on fourfold degenerate sites in {\em D.\ melanogaster}: Indeed, synonymous sites seem to be under varying selection strength including strong purifying selection. Generally, only directional selection is considered in the context of codon usage bias instead of balancing (or other forms of quadratic) selection. This follows \citet{Li87} and \citet{Bulm91}, who argued that mutation bias and opposing linear selection determine codon usage bias,  rather than balancing selection as \citet{Kimu81} postulated. Nevertheless, dominance and other non-additive effects could also contribute.

For {\em{D.\ melanogaster}} and {\em{D.\ simulans}}, the ratio of nucleotides $[AT]$ to nucleotides $[CG]$ is approximately $2:1$ in short autosomal introns, which likely reflects mutation bias \citep{Clem12a}. In fourfold degenerate sites of {\em{D.\ simulans}}, however, the ratio is approximately $1:2$. This likely reflects the joint action of mutation bias and directional selection \citep{Clem12b}. One can then define a polymorphism ratio that can be used as a proxy for directional selection as it correlates well with divergence measures \citep{Machado19, Lawrie20}. In populations of {\em{D.\ simulans}} (which are generally not too far from mutation-selection-drift equilibrium),  directional selection has a strength of approximately $\gamma=1.39$ favoring $C$ and $G$ nucleotides that compensates for the mutation bias in fourfold degenerate sites \citep{Vogl15,Jack17}. 

Usually, silent vs.\ replacement amino acid substitutions are compared with the McDonald-Kreitman test. Scenarios like the above suggest extending the approach to comparing short introns (as a neutral reference) with fourfold degenerate sites (which are under weak directional selection). Selection on the latter is so weak that reversible models must be considered. In this article, we derive equilibrium substitution rates (Sec.~\ref{sec:sub_rates}) and heterozygosities (Sec.~\ref{section:het}) for a boundary-mutation-(directional) selection-drift model. This makes it possible to go beyond testing for deviation from neutrality and also infer the strength of selection causing this change.

The layout of the article is as follows:
Sec.~\ref{section:historical_intro} provides a review of the boundary-mutation Moran model. Sec.~\ref{section:model} introduces the extension to linear and quadratic selection. We will see that considering a finite number of sites subject to a biallelic, reversible mutation scheme (with only one mutation segregating at a time) enables the derivation of an exact equilibrium distribution in Sec.~\ref{section:exact_stdistr}. We also provide a convenient approximation in Sec.~\ref{section:approx_stdistr}. We further calculate various statistics including a measure for expected heterozygosity that relates to both \citet{Kimu69a} and \citet{McVe99}, as well as the Ewens-Watterson estimator \citep{Ewen72,Ewen74,Watt75} in Sec.~\ref{section:het}, and simple formulae for substitution rates that relate to both the substitution rate and evolutionary rate of Kimura in Sec.~\ref{sec:sub_rates}. We use these estimators to infer the selection strength acting against mutation bias in fourfold degenerate sites of {\em Drosophila simulans} within the McDonald-Kreitman framework in Sec.~\ref{Sec:Droso_intro}.

\section{The boundary-mutation Moran model and diffusion approximations}
\label{section:historical_intro}

\subsection{Conceptual introduction to the boundary-mutation Moran model}
\label{section:historical_intro_pt1}

The Moran model was introduced as a model for genetic drift \citep{Mora58a, Mora58b, Mora62}. It assumes a monoecious, haploid population of $N$ individuals with alleles of a focal and non-focal type. At each step, a randomly chosen individual is replaced by the offspring of another randomly chosen individual. This system induces a tridiagonal transition matrix with absorbing states $0$ and $N$. With mutation, the process can always 'escape' the boundary states as well as drift into them, so the boundaries formally become partially reflecting \citep[][chapt.~2.2]{Karl75}. With the decoupled Moran model \citep{Baak08}, mutation, directional selection, and drift are parameterized as separate processes. All the above Moran models are equivalent to finite state birth-death processes with appropriate boundary conditions \citep[][chapt.~4]{Karl75}.

With pure drift and mutation, \ie\ without selection, and with a constant population size $N$ the equilibrium distribution of a sample (without replacement) of size $M$ (with $M\leq N$) is a beta-binomial distribution and the probability of obtaining a certain number of focal alleles $z$ in a biallelic setting becomes:

\begin{equation}\label{eq:beta_bin}
\begin{split}
  \Pr(Z=z\given M, \theta, \beta)&=\binom{M}{z} \frac{\Gamma(\theta)}{\Gamma(M+\theta)}\frac{\Gamma(z+\beta\theta)\Gamma(M-z+(1-\beta)\theta)}{\Gamma(\beta\theta)\Gamma((1-\beta)\theta)}\,,  
\end{split}
\end{equation}

where the scaled rate $\beta\theta$ is the mutation rate towards the focal allele (with $\theta$ denoting the mutation rate, $\beta$ the bias) and $(1-\beta)\theta$ the mutation rate away from it.

As with the Wright-Fisher model \citep[][chapter~4]{Ewen04}, Kolmogorov forward and backward equations can be derived with the Moran model; with the decoupled mutation-drift Moran model using only the first and second symmetric derivatives \citep{Berg17}. This is particularly relevant because standard population genetic results were derived by Kimura using the Kolmogorov backward equation, such as formulas for fixation probabilities \citep{Kimu62} and heterozygosity \citep{Kimu69a}. Inference based on the decoupled Moran model thus converges to these classic diffusion results \citep{Ethe09}. In particular, the beta-binomial is also the distribution of a sample of size $M$ from the population in the diffusion limit.

The boundary-mutation Moran model with mutation bias has thus far been studied for the case of either neutral evolution or linear selection \citep{Vogl12,Vogl15,Berg17}. The neutral boundary-mutation Moran model \citep{Vogl12} was originally introduced as a simplified decoupled Moran model, with the additional assumption that overall scaled mutation rates $\theta$ are sufficiently small such that mutations only occur at the monomorphic boundaries. The interior transitions of polymorphic sites are due to drift (or selection and drift). It is straightforward to derive the equilibrium distribution of the neutral equilibrium boundary-mutation Moran model \citep{Vogl12,Vogl15}.

Expanding the beta-binomial distribution above with a first order Taylor series in $\theta$ results in a distribution identical to that of a sample from the boundary-mutation Moran model \citep{Vogl14b}. Simulations show that this approximation holds well if the expected equilibrium heterozygosity $2\beta(1-\beta)\theta<0.025$ \citep{Vogl12}, where $\beta$ is the mutation bias towards the focal allele. In protein coding genes of eukaryotes the expected heterozygosity, which is approximately the scaled mutation rate, has been shown to be approximately $10^{-2}$ or less \citep{Lynch16}.

\subsection{Formal introduction to the boundary-mutation Moran model}
\label{section:historical_intro_pt2}

Consider a phenotypic trait in a population of small to moderate size $N$ that is influenced by $K$ sites, indexed by $k$ with $1\leq k \leq K$. Each site is assumed to be biallelic with one allele coded as $1$ and the other as $0$, so there are $2^K$ possible allelic combinations in total. Per generation (a generation corresponds to $N$ birth-death Moran events) and per site, a mutation from allele $0$ to allele $1$ occurs at a scaled rate $\beta\theta$; a mutation in the reverse direction occurs at a scaled rate $(1-\beta)\theta$. The effect of exchanging allele $0$ with allele $1$ is a proportional increase in fitness of the phenotype. Thus there are $K+1$ fitness states.

Assume that each site fixes independently. This assumption is approximately valid (i) if the scaled recombination rate is much larger than the scaled mutation rate, or (ii), in the case of very low effective recombination rates, if scaled mutation rates are so small that only one site segregates in the population at a time. Importantly, the assumption of independence usually holds for the small scaled mutation rates relevant for the boundary-mutation Moran model \citep{Vogl12}.

Now we can formally define the neutral boundary-mutation Moran model: Let $x(t)$ denote the relative frequency (or proportion) of allele $1$ at a focal locus at time $t$. In the interior, {\ie} for $x(t)=\frac{i}N$ with $1\leq i \leq N$-1, the transition probabilities from $t$ to $t+1$ are:
\begin{equation}\label{eq:transition_decoupled_Moran}
\begin{cases}
    \Pr(x(t+1)=\frac{i-1}N\given x(t)=\frac{i}{N})&=\frac{i(N-i)}{N^2}\\
    \Pr(x(t+1)=\frac{i}{N}\given x(t)=\frac{i}{N})&=1-2\frac{i(N-i)}{N^2}\\
    \Pr(x(t+1)=\frac{i+1}N\given x(t)=\frac{i}{N})&=\frac{i(N-i)}{N^2}\,.
\end{cases}
\end{equation}
At the boundary with $i=0$, we have:
\begin{equation}\label{eq:transition_decoupled_Moran_0}
\begin{cases}
    \Pr(x(t+1)=0\given x(t)=0)&=1-\beta\tfrac{\theta}{N}\frac{1}{1-\beta{\theta} H_{N-1}}\\
    \Pr(x(t+1)=\frac{1}{N}\given x(t)=0)&=\beta\tfrac{\theta}{N}\frac{1}{1-\beta{\theta} H_{N-1}}\,,
\end{cases}
\end{equation}
where $H_{n}=\sum_i^n \frac{1}{i}$ is the harmonic number. The normalizing term $\frac{1}{1-\beta{\theta} H_{N-1}}$ ensures that, in equilibrium, mutations enter the polymorphic region at an identical average rate $\beta\tfrac{\theta}{N}$ per Moran drift event, irrespective of $N$. At the boundary $i=N$, we have analogously:
\begin{equation}\label{eq:transition_decoupled_Moran_N}
\begin{cases}
    \Pr(x(t+1)=1\given x(t)=1)&=1-(1-\beta)\tfrac{\theta}{N}\frac{1}{1-(1-\beta){\theta} H_{N-1}}\\
    \Pr(x(t+1)=\frac{N-1}{N}\given x(t)=1)&=(1-\beta)\tfrac{\theta}{N}\frac{1}{1-(1-\beta){\theta}H_{N-1}}\,.
\end{cases}
\end{equation}
Since a generation corresponds to $N$ Moran events, the mutation rates must be multiplied by $N$ to obtain the mutation rate per generation. 

Note that with the general mutation model, mutations mainly arise from close to the boundaries when mutation rates are low. With the boundary mutation model, mutations arise exclusively from the boundaries. The terms normalizing the mutation rates at the boundaries in Eq.~(\ref{eq:transition_decoupled_Moran_0}) and Eq.~(\ref{eq:transition_decoupled_Moran_N}) compensate for this difference. 

The equilibrium distribution of the proportion of alleles $\mathbf{X}$ at each locus is then \citep{Vogl14b}:
\begin{equation}\label{eq:equilibrium_nosel}
\mathbf{\pi}=\Pr(\mathbf{X} = \frac{i}{N} \mid N,\beta,\theta)=
\begin{cases}
    \displaystyle (1-\beta)(1-\beta\theta H_{N-1}) & i=0; \\
    \displaystyle \beta(1-\beta)\theta\frac{N}{i(N-i)} & 1\leq i \leq N-1;\\
    \displaystyle \beta(1-(1-\beta)\theta H_{N-1}) & i=N.
\end{cases}
\end{equation}
From the equilibrium distribution it follows that the distribution of a sample of size $M$ taken without replacement is independent of $N$. 

An irreducible, positive recurrent, aperiodic Markov chain with a tridiagonal transition matrix is reversible. Thus the distribution Eq.~(\ref{eq:equilibrium_nosel}) can be shown to be the equilibrium distribution, as it fulfills detailed balance. Note that the boundary-mutation Moran model is also a birth-death process with partially reflecting boundaries, as is the general mutation Moran model. Thus the stationary distribution can also be derived using the theory of a finite birth-death process  \citep[][chapt.~4.6]{Karl75}:
\begin{equation}
    \begin{split}
        &\pi_i=\pi_0\frac{\beta\theta}{1-\beta\theta H_{N-1}}\frac{N}{i(N-i)}\qquad\text{for $1\leq i \leq N-1$,}\\
        &\pi_N=\pi_0\frac{\beta}{1-\beta}\frac{1-(1-\beta)\theta H_{N-1}}{1-\beta\theta H_{N-1}}\,,
    \end{split}
\end{equation}
with
\begin{equation}
            \pi_0=\frac{1}{1+\sum_{i=1}^{N}\pi_i}\,.
\end{equation}
Equivalence to the stationary distribution~(Eq.~\ref{eq:equilibrium_nosel}) follows immediately. The symmetry between the boundary terms $\pi_0$ and $\pi_N$ and the difference of both from the polymorphic terms are, however, not as readily apparent as in Eq.~(\ref{eq:equilibrium_nosel}).

\subsection{Population size and polymorphism limits}\label{Section:Popsizelimit_no_sel}

Note that for large $N$ either $\Pr(\mathbf{X}=0)$, $\Pr(\mathbf{X}=1)$, or both become negative. This naturally imposes a limit on the population size for the boundary-mutation Moran model, making it valid only for small to moderate population sizes depending on $\theta$. More precisely: Assume without loss of generality that $\beta<0.5$ (which can be achieved by convenient labelling of alleles). Then $\beta-\beta(1-\beta)\theta H_{N-1}>0$ must hold. For large $N$, $H_{N-1}\approx \log(N-1)$ and therefore $N<e^{\tfrac{1}{(1-\beta)\theta}}$ must hold. Recall that the upper limit of validity of the boundary-mutation model as an approximation to the general mutation model is approximately $2\beta(1-\beta)\theta=0.025$ \citep{Vogl12}, and that mutation bias is usually not extreme. Therefore, we determine $N<e^{10}$ as an approximate upper bound for the population size. This is larger than most effective population sizes and therefore hardly a practical limitation. Note that with $N$ close to this limit, probability mass is focused mainly in the polymorphic region just as with the general Moran model, because the proportionate increase in mutation probabilities from the boundaries in Eq.~(\ref{eq:transition_decoupled_Moran_N}) compensates for the absence of mutations in the polymorphic region. We will return to these considerations in Sec.~\ref{Section:Popsizelimit_with_sel}.

\begin{figure}[!ht]
    \begin{center}
    \includegraphics[scale=0.40]{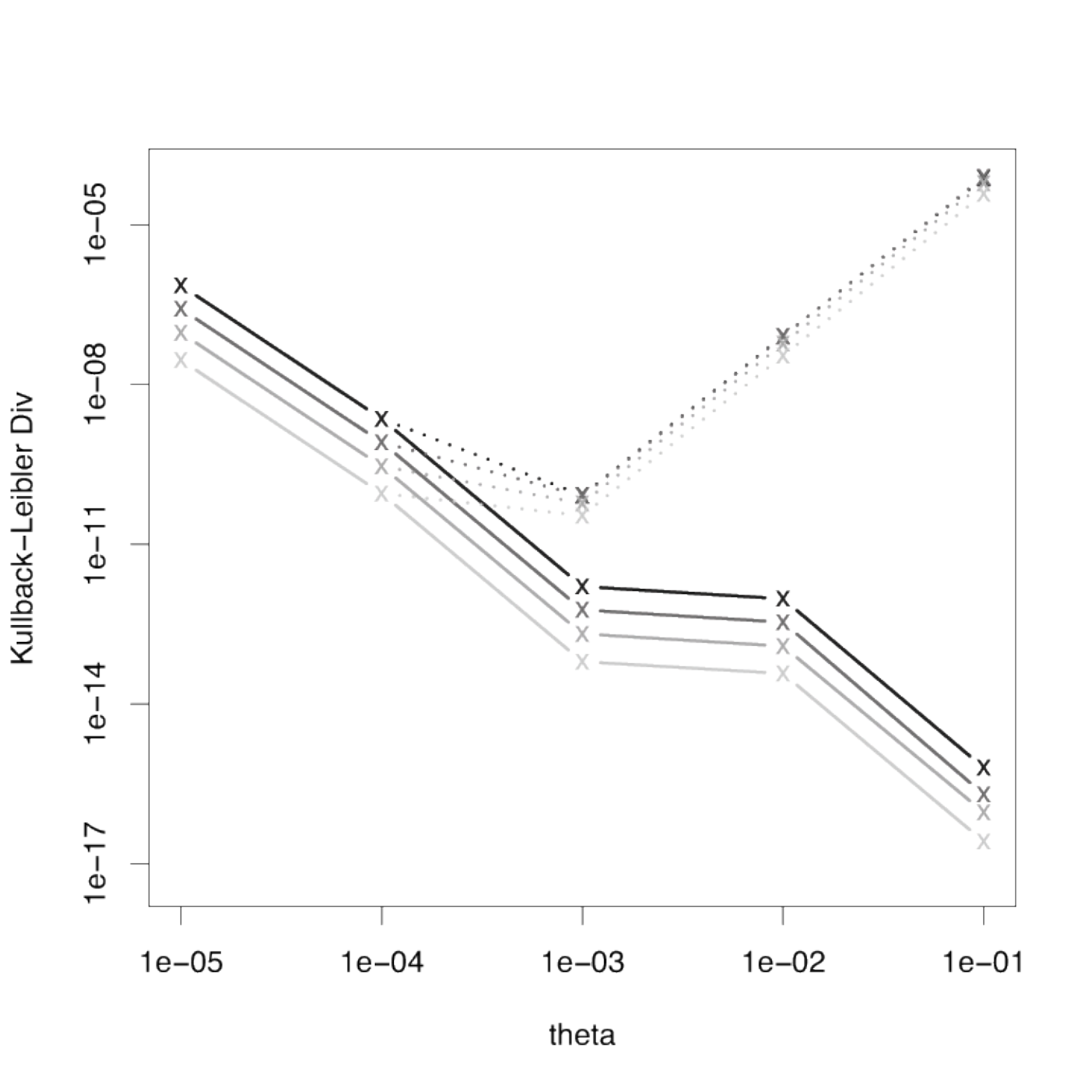}
    \end{center}
    \begin{center}
    \caption{Kullback-Leibler divergence: We start with $5$ neutrally evolving populations of size $N=1000$ with mutation bias $\beta=1/3$ and mutation rates $\theta=(0.00001,0.0001,0.001,0.01,0.1)$ respectively. We model them as evolving according to a general mutation Moran model which is here identical to Wright's equilibrium. For each $\theta$, we sample from these populations with replacement to obtain population sample of sizes $M=(3,10,30,100)$. For each $\theta$, we then calculate the equilibrium distributions of a general and a boundary-mutation Moran model with population sizes equal to the sample sizes and compare to the corresponding downsampled equilibrium distribution via the Kullback-Leibler divergence. This is defined as $D_{KL}=-\sum_i p_i\log(\tfrac{q_i}{p_i})$, where $q_i$ are the individual allele frequencies in the downsampled equilibrium and $p_i$ are those of the respective model used for the sample. Above, we show this divergence (y-axis) for varying $\theta$ (x-axis). The equilibrium distributions of samples modelled by the general mutation Moran model vs.\ the downsampled population are represented by solid lines, and the samples modelled by a boundary-mutation Moran model vs.\ the downsampled population by dotted lines. The shading of both line types becomes lighter for increasing sample size.}
    \label{fig:KL1}
    \end{center}
\end{figure}

Let us now address the upper limit of polymorphism permitted in a boundary-mutation Moran model that we referred to in the previous paragraph. 
Below, we show the Kullback-Leibler (KL) divergence between two distributions: i) samples taken without replacement from the beta-binomial distribution that corresponds with Wight's equilibrium distribution \citep{Wrig31} (these samples are also beta-binomially distributed) and ii) the equilibrium distributions of the general mutation Moran model and the boundary-mutation Moran model respectively, both with $N$ always of the same size as the sample drawn from the population, for varying mutation rates (Fig.~\ref{fig:KL1}). Note that in the case of the general mutation model, a sample from the stationary distribution also conforms to a beta-binomial compound distribution, since the stationary distribution is the beta distribution \citep{Wrig31}. Hence the beta-binomial distributions from i) and ii) are identical in this. Thus the KL divergences should be zero for the general mutation Moran model and the observed deviations are caused by numerical errors. The divergence estimates of the general mutation Moran model and the boundary-mutation Moran model only start to differ noticeably for $\theta>0.01$ and the difference remains small even for $\theta=0.1$. This speaks for the approximation accuracy of the boundary-mutation Moran model. Again, we will return to this topic in Sec.~\ref{Section:Popsizelimit_with_sel}.

\newpage

\section{Moran model with biased mutation, linear, and quadratic selection}
\label{section:model}

\subsection{Selection coefficients}
\label{section:sel_coefs}

Note that with a strictly haploid model dominance and over- and underdominance are impossible. A diploid selection model allowing for these effects involves two alleles that are lost or gained when a diploid individual competes against another diploid individual. With two alleles lost, the transition matrix could no longer be tridiagonal. In order to model diploids in a haploid framework with a tridiagonal matrix, we use a similar argument for obtaining selection coefficients to \citet{Muir09}. Consider a population in Hardy-Weinberg equilibrium. The focal allele partners up with an allele randomly drawn from the population to obtain its fitness; it competes with another allele, which also obtains its fitness by partnering up with a further allele randomly drawn from the population. We define $B_1$ as determining the strength of first order and $B_2$ the strength of second order selection (so if $B_2=0$, the fitness effects are purely additive). Then the relative fitnesses for a diploid individual with $0$, $1$, and $2$ alleles of the focal type are given by $1$, $1+\tfrac{B_1}{2N}+\tfrac{B_2}{2N}$, and $1+2\tfrac{B_1}{2N}$. The relative differences in fitness of two (ordered) genotypes are given Table~(\ref{Table:Sel_coeff}), where the competing alleles are in bold. The two columns on the right correspond to the ordered genotypes containing the focal allele, the two on the left to the ordered genotypes involving the competitor allele; and analogously for the rows. 
\begin{table}[h!]
\caption{Fitnesses of two genotypes}
\label{Table:Sel_coeff}
\begin{tabular}{c|cccc}
  & $\mathbf{0}0$ & $\mathbf{0}1$ & $\mathbf{1}0$ & $\mathbf{1}1$ \\
\hline
$\mathbf{0}0$ & $0$ &
$\tfrac{B_1}{2N}+\tfrac{B_2}{2N}$ & $\tfrac{B_1}{2N}+\tfrac{B_2}{2N}$ & $\tfrac{2B_1}{2N}$ \\
$\mathbf{0}1$ & $-\tfrac{B_1}{2N}-\tfrac{B_2}{2N}$ & $0$ & $0$ & $\tfrac{B_1}{2N}-\tfrac{B_2}{2N}$\\
$\mathbf{1}0$ & $-\tfrac{B_1}{2N}-\tfrac{B_2}{2N}$ & $0$ & $0$ & $\tfrac{B_1}{2N}-\tfrac{B_2}{2N}$ \\
$\mathbf{1}1$ & $-\tfrac{2B_1}{2N}$ & $-\tfrac{B_1}{2N}+\tfrac{B_2}{2N}$ & $-\tfrac{B_1}{2N}+\tfrac{B_2}{2N}$ & $0$
\end{tabular}
\end{table}
The focal allele replaces its competitor according to their marginal fitness difference, whereas the two partner alleles remain unaffected. In other words, the probability of a selective change in the allele frequency from $i$ to $i+1$ per Moran event is:

\begin{equation}
    \begin{split}
        \frac{(N-i)i}{N^2}s_{i\rightarrow i+1}&=\frac{(N-i)i}{2N^5}\bigg((N-i)^2(B_1+B_2)+i^2(-B_1+B_2)\\
        &\qquad\qquad\qquad\qquad+2i(N-i)B_1+2i^2(B_1-B_2)\bigg)\\
        s_{i\rightarrow i+1}
        &=\tfrac{B_1}{2N}-\tfrac{B_2(2i-N)}{2N^2}\,,
    \end{split}
\end{equation}
where $s_{i\rightarrow i+1}$ is the selection coefficient. The selection coefficient in the reverse direction is analogously:
\begin{equation}
    s_{i\rightarrow i-1}=-\frac{B_1}{2N}+\frac{B_2(2i-N)}{2N^2}\,.
\end{equation}

While the Kolmogorov forward (\ie\  diffusion) and backward equations  generally can be derived from Wright-Fisher \citep[][chapt.~4]{Ewen04} and Moran models, such a derivation requires only the definition of the first and second symmetric derivatives for the decoupled mutation-drift Moran model \citep[][Appendix~7.1]{Berg17}. Following this procedure, one easily sees that the diffusion approximation of a boundary-mutation Moran model including the selection coefficients above corresponds to Kimura's diffusion approach \citep{Kimu81} except for a reversal of signs for the parameter $B_2$. This is because Kimura starts from a normally distributed phenotype, assumes a fitness function proportional to a normal distribution, and then considers the response of a biallelic locus influencing the trait under stabilizing phenotypic selection. Near the fitness optimum there are two possible scenarios: i) The population itself is close to the fitness optimum, resulting in linear selection towards the optimum; ii) the population is right at the optimum, resulting in underdominance. Thus, while in our case a positive $B_2$ corresponds to overdominant selection, in Kimura's case it corresponds to underdominant selection.

\subsection{Exact stationary distribution}
\label{section:exact_stdistr}

Using the selection coefficients derived in the previous subsection, we get the exact interior transition probabilities:
\begin{equation}\label{eq:transition_decoupled_Moran_linear_sel}
\begin{cases}
    \Pr(x(t+1)=\frac{i-1}N\given x(t)=\frac{i}{N})&=\big(1-\tfrac{B_1}{2N}+\tfrac{B_2(2i-N)}{2N^2}\big)\frac{i(N-i)}{N^2}\\
    \Pr(x(t+1)=\frac{i}{N}\given x(t)=\frac{i}{N})&=1-2\frac{i(N-i)}{N^2}\\
    \Pr(x(t+1)=\frac{i+1}N\given x(t)=\frac{i}{N})&=\big(1+\tfrac{B_1}{2N}-\tfrac{B_2(2i-N)}{2N^2}\big)\frac{i(N-i)}{N^2}\,.
\end{cases}
\end{equation}
Note that this transition matrix deviates from the one used earlier for only linear selection \citep{Vogl12,Vogl15}, but converges to the same diffusion limit. At the boundary $i=0$, we have the boundary transitions:
\begin{equation}\label{eq:transition_decoupled_Moran_linear_sel_0}
\begin{cases}
    \Pr(x(t+1)=0\given x(t)=0)&=1-\beta\tfrac{\theta}{N}\big(1+\tfrac{B_1}{2N}+\frac{B_2}{2N}\big)C_0\\    
    \Pr(x(t+1)=\frac{1}{N}\given x(t)=0)&=\beta\tfrac{\theta}{N}\big(1+\tfrac{B_1}{2N}+\frac{B_2}{2N}\big)C_0\,,
\end{cases}
\end{equation}
with
\begin{equation}
   C_0=\bigg(1-\beta\theta\sum_{i=1}^{N-1} \frac{1}{i}\,R_i\bigg)^{-1}\,.
\end{equation}
 where 
 \begin{equation}
    \begin{split}
 R_i&=\bigg(\frac{\prod_{j=0}^{i-1}(1+\tfrac{B_1}{2N}-\tfrac{B_2(2i-N)}{2N^2})}{\prod_{j=1}^{i}(1-\tfrac{B_1}{2N}+\tfrac{B_2(2i-N)}{2N^2})}\bigg)\,.
 \end{split}
\end{equation}
 
Set
\begin{equation}
    \begin{split}
        R_N&=\frac{\prod_{i=0}^{N-1}(1+\tfrac{B_1}{2N}-\frac{B_2(2i-N)}{2N^2})}{\prod_{i=0}^{N-1}(1-\tfrac{B_1}{2N}+\frac{B_2(2i-N)}{2N^2})}\,.
    \end{split}
\end{equation}
At the boundary $i=N$, we then have:
\begin{equation}\label{eq:transition_decoupled_Moran_linear_sel_N}
\begin{cases}
    \Pr(x(t+1)=1\given x(t)=1)&=1-(1-\beta)\tfrac{\theta}{N}\big(1-\tfrac{B_1}{2N}-\frac{B_2}{2N}\big)C_1\\
    \Pr(x(t+1)=\frac{N-1}N\given x(t)=1)&=(1-\beta)\tfrac{\theta}{N}\big(1-\tfrac{B_1}{2N}-\frac{B_2}{2N}\big) C_1\,,
\end{cases}
\end{equation}
with
\begin{equation}
    C_1=\bigg(1-(1-\beta)\theta R_N^{-1} \sum_{i=1}^{N-1} \frac{1}{N-i}\, R_i\bigg)^{-1}\,.
\end{equation}

Set
\begin{equation}
    \omega =\frac{\beta (1-\beta)\theta }{(1-\beta)+ \beta R_N}
\end{equation} 
and 
\begin{equation}
    \varphi =\frac{\beta R_N}{(1-\beta)+\beta R_N}\,.
\end{equation}
It follows that the exact equilibrium distribution for a boundary-mutation Moran model with both linear and quadratic selection as well as biased mutation can be written as:
\begin{equation}\label{eq:equilibrium_linsel}
\mathbf{\pi}=\Pr(\mathbf{X} = \tfrac{i}{N} \given B_1,B_2,\varphi ,\omega)=
    \begin{cases}
    \displaystyle (1-\varphi )-\omega \sum_{i=1}^{N-1}\frac{1}{i}R_i & i=0; \\
    \\
    \displaystyle 
    \omega \frac{N}{i(N-i)} R_i & 1\leq i\leq N-1;\\
    \\
    \displaystyle  \varphi -\omega \sum_{i=1}^{N-1}\frac{1}{N-i}R_i& i=N\,.
\end{cases}
\end{equation}

We recall that because the transition matrix is tridiagonal, detailed balance must hold for nearest neighbours in equilibrium. In the interior, so for $1\leq i+1 \leq N-1$, the flow balances, as the flow from the state $\mathbf{X}=\frac{i+1}{N}$ to $\mathbf{X}=\frac{i}{N}$ is:
\begin{equation}
\begin{split}
    &\Pr\big(x(t+1)=\tfrac{i-1}N\given x(t)=\tfrac{i}{N}\big) \Pr\big(\mathbf{X}=\tfrac{i}{N}\big)\\
    &=\bigg(1+\tfrac{B_1}{2N}-B_2\frac{2i-N}{2N^2}\bigg)\frac{i(N-i)}{N^2}\omega  \frac{\prod_{j=0}^{i-1}(1+\frac{B_1}{2N}-B_2\frac{2j-N}{2N^2})}{\prod_{j=1}^{i}(1-\frac{B_1}{2N}+B_2\frac{2j-N}{2N^2})}\frac{N}{i(N-i)}\\
    &=\frac{\omega}{N}\frac{\prod_{j=0}^{i}(1+\frac{B_1}{2N}-B_2\frac{2j-N}{2N^2})}{\prod_{j=1}^{i}(1-\frac{B_1}{2N}+B_2\frac{2j-N}{2N^2})}\,,
\end{split}
\end{equation}
and that in the reverse direction:
\begin{equation}
\begin{split}
    &\Pr\big(x(t+1)=\tfrac{i}N\given x(t)=\tfrac{i+1}{N}\big) \Pr\big(\mathbf{X}=\tfrac{i+1}{N}\big)\\
    &=\bigg(1-\tfrac{B_1}{2N}+B_2\frac{2(i+1)-N}{2N^2}\bigg)\frac{(i+1)(N-i-1)}{N^2}\omega  \\
    &\qquad \times\frac{\prod_{j=0}^{(i+1)-1}(1+\frac{B_1}{2N}-B_2\frac{2j-N}{2N^2})}{\prod_{j=1}^{(i+1)}(1-\frac{B_1}{2N}+B_2\frac{2j-N}{2N^2})}\frac{N}{(i+1)(N-i-1)}\\
    &=\frac{\omega}{N}\frac{\prod_{j=0}^{i}(1+\frac{B_1}{2N}-B_2\frac{2j-N}{2N^2})}{\prod_{j=1}^{i}(1-\frac{B_1}{2N}+B_2\frac{2j-N}{2N^2})}\,.
    \end{split}
\end{equation}
At the boundary $i=0$, the flow also balances, as the flow from $\mathbf{X}=0$ to $\mathbf{X}=\tfrac{1}{N}$ is:
\begin{equation}
\begin{split}
    &\Pr\big(x(t+1)=\tfrac{1}N\given x(t)=0\big)\Pr\big(\mathbf{X}=0\big)\\ &=\beta\frac{\theta}{N}\big(1+\tfrac{B_1}{2N}+B_2\frac{N}{2N^2}\big)C_0
    \bigg((1-\varphi)-\omega \sum_{i=1}^{N-1}\frac{1}{i} \frac{\prod_{j=0}^{i-1}(1+\frac{B_1}{2N}-B_2\frac{2j-N}{2N^2})}{\prod_{j=1}^{i}(1-\frac{B_1}{2N}+B_2\frac{2j-N}{2N^2})}\bigg)\\ 
    &=\beta\frac{\theta}{N}\big(1+\tfrac{B_1}{2N}+B_2\frac{N}{2N^2}\big)C_0
    (1-\varphi )\frac{1}{C_0}\\
    &=\frac{\omega}{N}\big(1+\tfrac{B_1}{2N}+B_2\frac{N}{2N^2}\big) \,,
\end{split}
\end{equation}
and that in the reverse direction:
\begin{equation}
\begin{split}
    &\Pr\big(x(t+1)=0\given x(t)=\tfrac{1}{N}\big) \Pr\big(\mathbf{X}=\tfrac{1}{N}\big)\\
    &=\big(1-\tfrac{B_1}{2N}+B_2\frac{2-N}{2N^2}\big)\frac{(N-1)}{N^2} 
    \omega  \bigg(\frac{1+\tfrac{B_1}{2N}+B_2\frac{N}{2N^2}}{1-\tfrac{B_1}{2N}+B_2\frac{2-N}{2N^2}}\bigg)\frac{N}{(N-1)}\\
    & = \frac{\omega}{N}\big(1+\tfrac{B_1}{2N}+B_2\frac{N}{2N^2}\big) \,.
\end{split}
\end{equation}
The other boundary $i=N$ follows analogously.

We have thus validated Eq.~(\ref{eq:equilibrium_linsel}) as the unique stationary distribution, as long as the boundary terms at $i=0$ and $i=N$ are positive. As any other Markov process with a tridiagonal transition matrix, the boundary-mutation Moran model with both linear and quadratic selection and biased mutation corresponds to a finite birth-death process. 

\subsection{Population size and polymorphism limits}\label{Section:Popsizelimit_with_sel}

\begin{figure}[!ht]
    \begin{center}
    \includegraphics[scale=0.45]{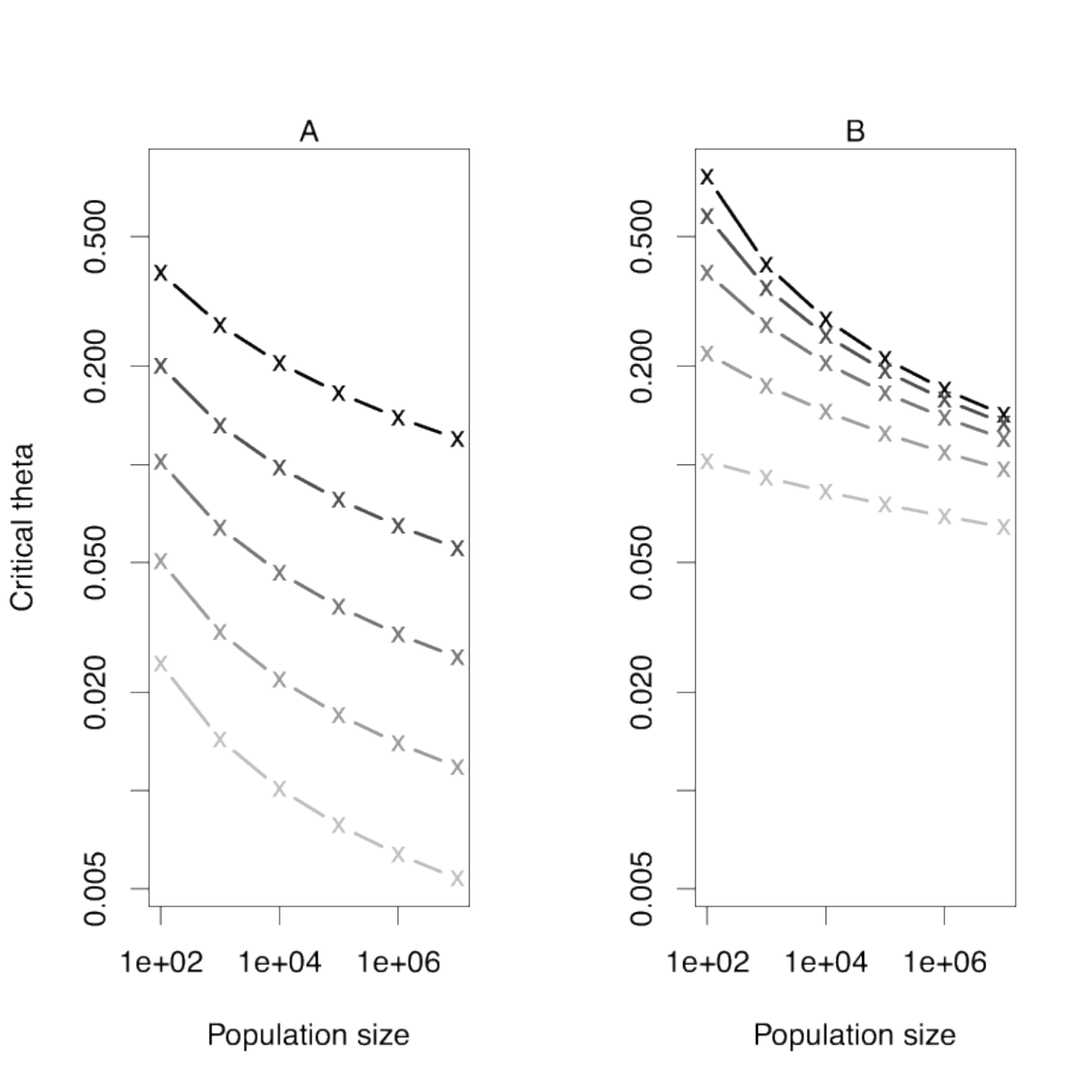}
    \end{center}
    \begin{center}
    \caption{For varying population size $N$ (x-axis), fixed mutation bias $\beta= 1/2$ (no bias), and varying selection strengths (A) $B_1=c(0,1,2,3,4)$, $B_2=0$ and (B) $B_1=0$, $B_2=c(-10,-5,0,5,10)$  respectively, we determine the maximum value of $\theta$ (y-axis) permitted by the boundary-mutation Moran model without the boundary terms becoming negative. The shade of the lines decreases with increasing positive value of selection strength.}
    \label{fig:CritTheta}
    \end{center}
\end{figure}

With selection, it is less straightforward to determine a closed form upper bound for the population size of the boundary-mutation Moran model, which we were able to do for the neutral case (Sec.~\ref{Section:Popsizelimit_no_sel}). Assuming no mutation bias, we evaluated the maximum mutation rate possible for a range of population sizes without incurring negative boundary terms (Fig.~\ref{fig:CritTheta}). We see that directional selection substantially reduces the strength of mutation rates that can be modelled in a population of a given size (Fig.~\ref{fig:CritTheta}A). However, even the combination of large populations and strong nearly-neutral directional selection is unlikely to invalidate the use the boundary-mutation Moran model in eukaryote systems \citet{Lynch16}. Quadratic selection impacts the critical combination of mutation rate and population size less severely (Fig.~\ref{fig:CritTheta}B).
\newpage

\newpage

\begin{figure}[!ht]
    \begin{center}
    \includegraphics[scale=0.45]{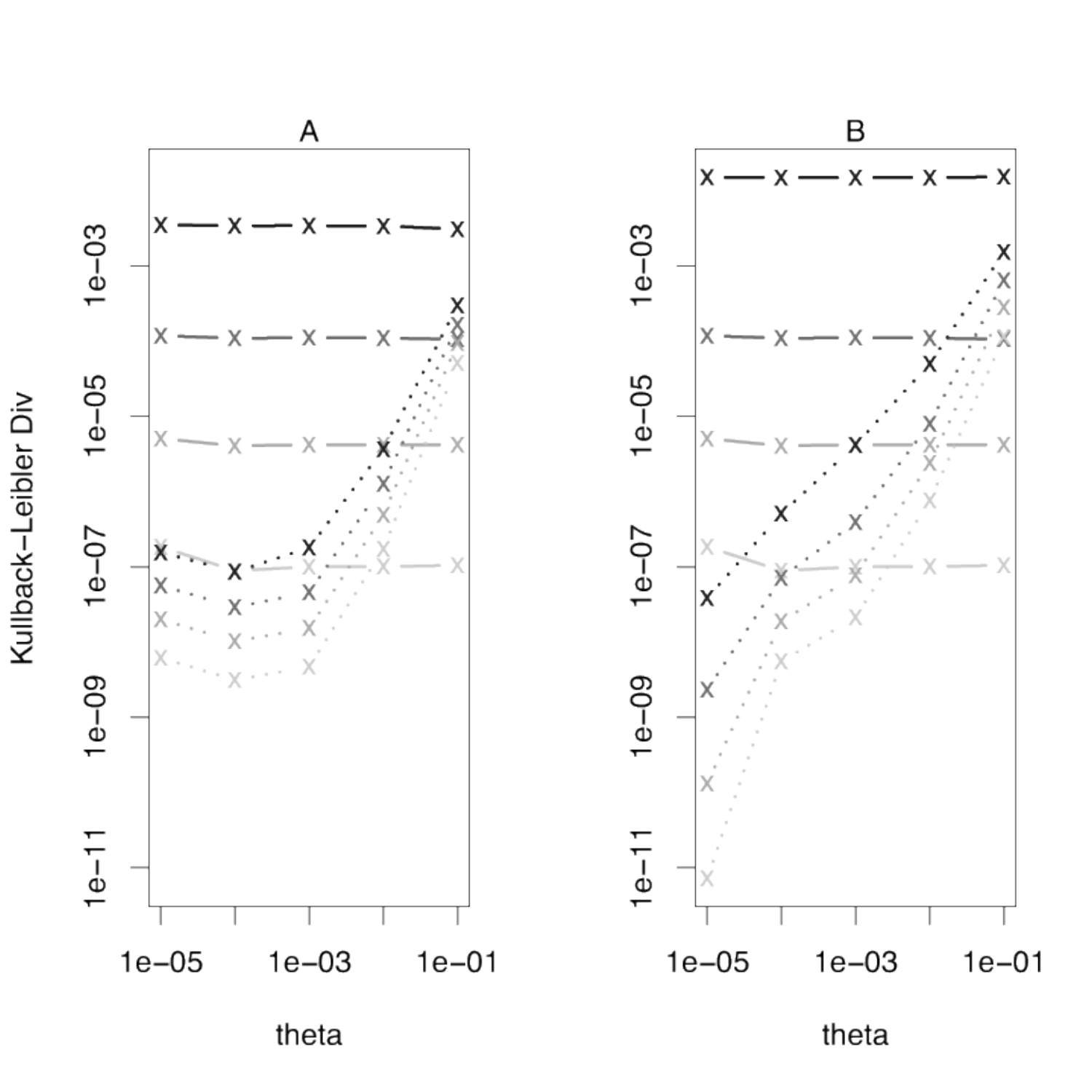}
    \end{center}
    \begin{center}
 \caption{Kullback Leibler divergence for directional selection: The stationary distribution of populations of size $N=1000$ with mutation bias $\beta=1/3$ and mutation rates $\theta=(0.00001,0.0001,0.001,0.01,0.1)$ respectively, is calculated for a general mutation Moran model with $B_1= 1$, $B_2= 0$  (A) and $B_1$= 3, $B_2= 0$ (B). We show the KL divergence (y-axis) for varying $\theta$ (x-axis) between the equilibrium distributions of the samples modelled by a general mutation Moran model vs.\ the downsampled population (solid line), and the samples modelled by a boundary-mutation Moran model vs.\ the downsampled population (dotted line) for sample sizes $M=(3,10,30,100)$ (the shading of both types of lines becomes lighter with increasing sample size). For details about the divergence measure see Fig.~\ref{fig:KL1}.}
\label{fig:KL2}
\end{center}
\end{figure}

\begin{figure}[!ht]
    \begin{center}
    \includegraphics[scale=0.45]{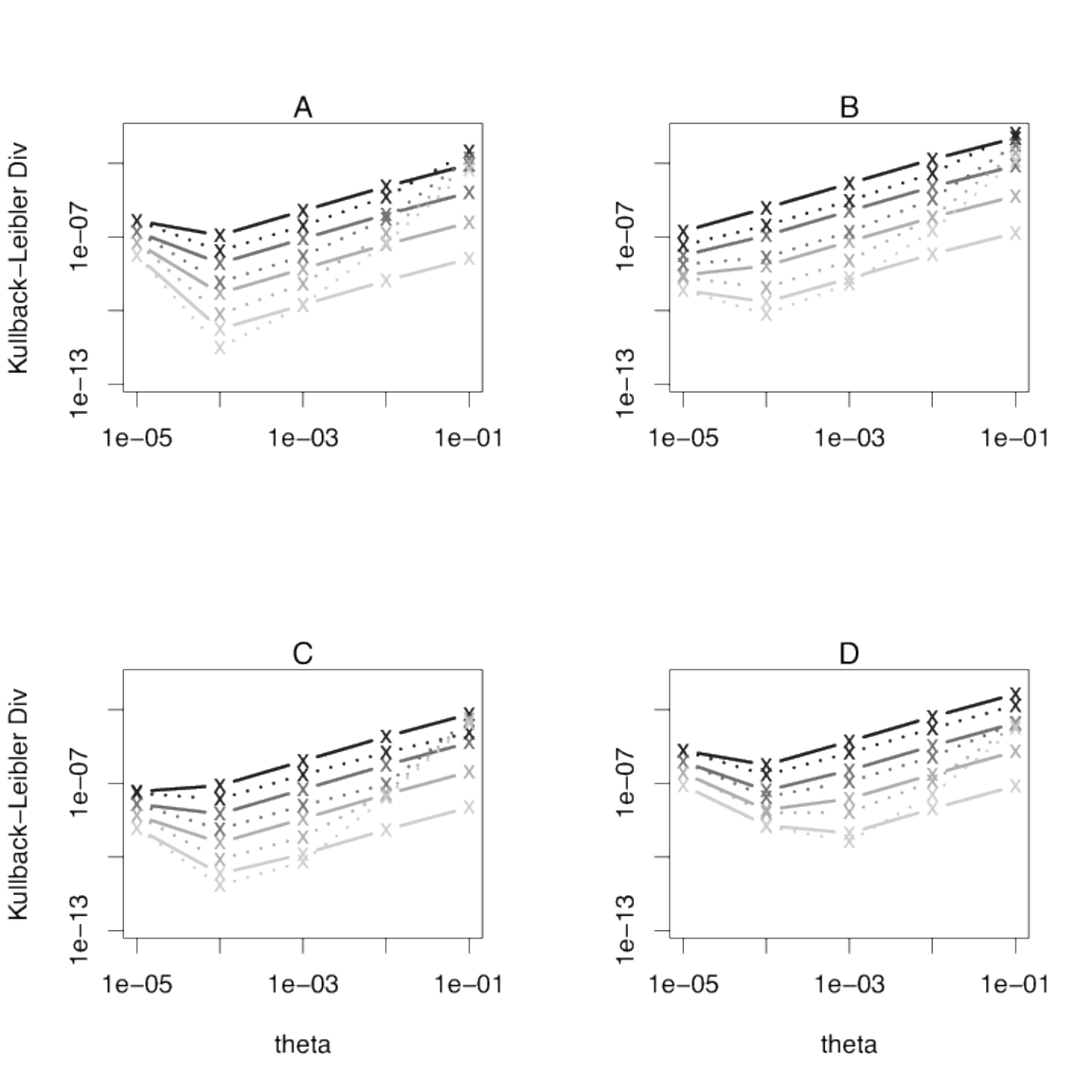}
    \end{center}
    \begin{center}
    \caption{Kullback-Leibler divergence for balancing selection: The stationary distribution of populations of size $N=1000$ with mutation bias $\beta=1/3$ and mutation rates $\theta=(0.00001,0.0001,0.001,0.01,0.1)$ respectively, is calculated for a general mutation Moran model with $B_1= 0$, $B_2= 1$  (A) and $B_1$= 0, $B_2= 3$ (B) and $B_1= 0$, $B_2= -1$  (C) and $B_1$= 0, $B_2= -3$ (D). We show the Kullback Leibler divergence (y-axis) for varying $\theta$ (x-axis) between the equilibrium distributions of the samples modelled by a general mutation Moran model vs. the downsampled population (solid line), and the samples modelled by a boundary-mutation Moran model vs. the downsampled population (dotted line) for sizes $M=(3,10,30,100)$ (shades of both types of lines decrease with increasing sample size). For details about the divergence measure see Fig.~\ref{fig:KL1}.}
    \label{fig:KL3}
    \end{center}
\end{figure}

We also evaluate the difference between modelling samples drawn from a larger population with a general mutation scheme using either a general mutation Moran model or a boundary-mutation Moran model by comparing Kullback-Leibler (KL) divergences as in Sec.~\ref{Section:Popsizelimit_no_sel}. This time, however, we include varying strengths of directional and quadratic selection: For a combination of directional selection and low mutation rates, the boundary-mutation Moran model with small sample sizes seems to approximate the large population (which can be thought of as close to the stationary distribution of the diffusion equation) better than the general mutation model with the same sample sizes (Fig.~\ref{fig:KL2}). This effect becomes more pronounced with greater selection strengths, and although it tapers off for increasing population sizes and increasing mutation rates, it does hold many reasonable parameter combinations. The qualitative difference between the divergence estimates obtained by modelling the samples with either a general or boundary-mutation Moran model in the presence of quadratic selection seem similar to the results for directional selection except perhaps for larger samples and lower negative values of quadratic selection (Fig.~\ref{fig:KL3}). Note, however, that the order of divergence is in the range of numerical errors, such that the equilibrium distributions of the models are nearly indistinguishable. These results suggest that working with the boundary-mutation Moran model is advantageous when forced to work with small sample sizes, either for numerical or other practical reasons.
\newpage

\newpage

\newpage

\begin{figure}[!ht]
    \begin{center}
    \includegraphics[scale=0.45]{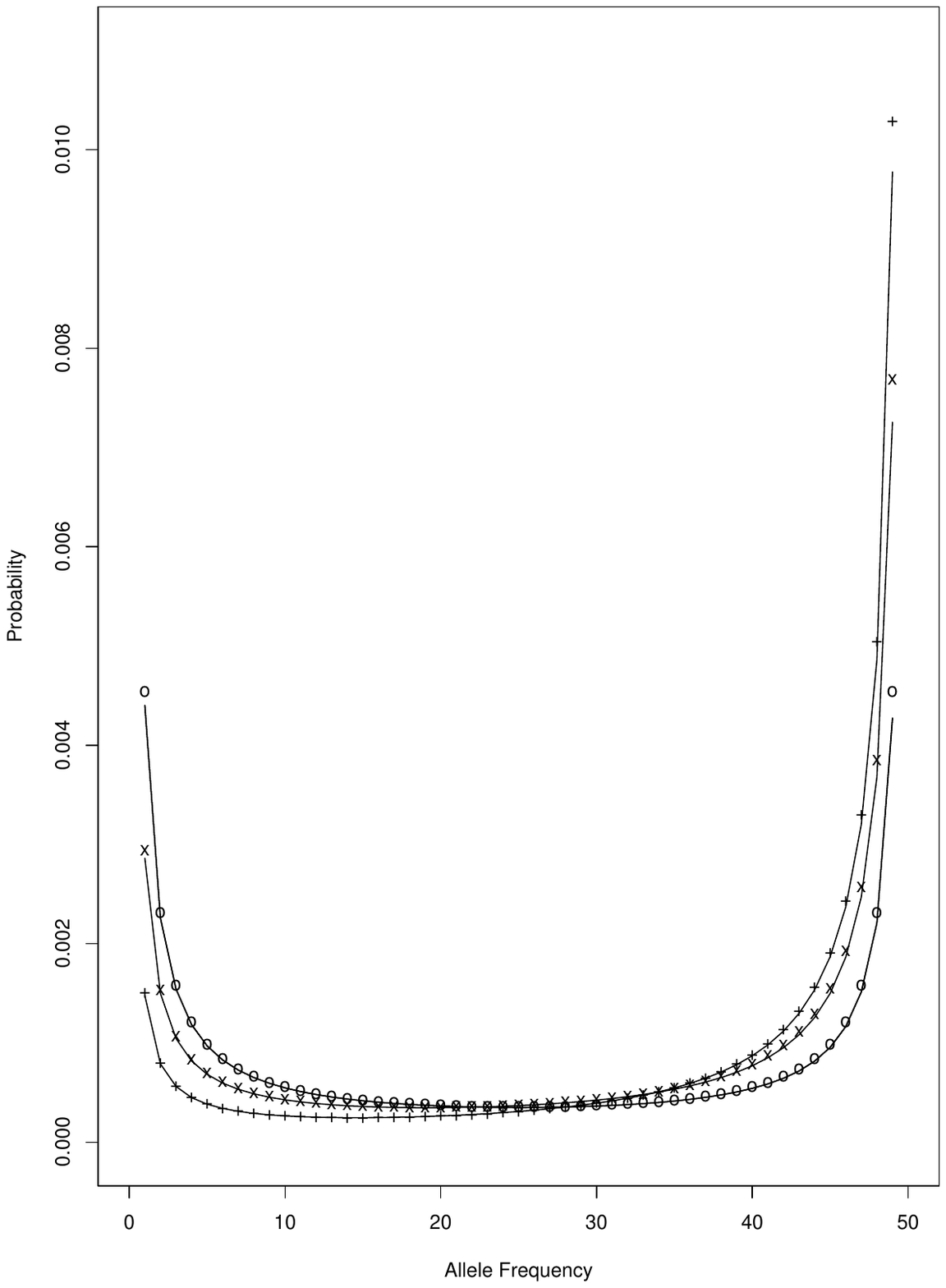}
    \end{center}
    \begin{center}
    \caption{The equilibrium probability (y-axis) of observing the focal allele within the polymorphic region at a certain frequency (x-axis) with Wright's equilibrium distribution (solid line) and the boundary-mutation Moran model (symbols: `o', `+', and `x' for $\gamma=(0,1,2)$, respectively). Parameters are: $\beta=1/3$, $\theta=0.02$, $N=50$, $B_1=(0,1,2)$, and $B_2=0$.}
    \label{fig:ContinuousDiscrete}
    \end{center}
\end{figure}

\newpage

\subsection{Approximate stationary distribution}
\label{section:approx_stdistr}

In this subsection, we find an exponential approximation to the exact boundary-mutation Moran model with biased mutation, and directional and quadratic selection. We will see that it has a simpler form that may be advantageous for implementation purposes and that it provides a more immediate comparison to classic diffusion results in Sec.~\ref{section:statistics}.

\subsubsection{Asymptotics of the drift and selection terms}
\label{app:asymptotics}

Let us examine the asymptotics of the terms for drift and selection: Assuming that $N$ is suitably large and $B_1$ and $B_2$ are at most of first order, the numerator can be approximated by:
\begin{equation}
\begin{split}
    f_i&=\prod_{j=0}^{i-1}\bigg(1+\tfrac{B_1}{2N}-B_2\tfrac{2j-N}{2N^2}\bigg)
    =\exp\bigg(\tfrac1{2N}\bigg(iB_1+B_2\tfrac{i(N-i+1)}{N}\bigg)\bigg)+\mathcal{O}(1/N^2)\,.
\end{split}
\end{equation}
Note that this is essentially a first order Taylor expansion of the exponential in reverse. From $e^s=(1+s)+\mathcal{O}(s^{2})$, we can see that the exponential reliably approximates the exact process for the large population sizes usually encountered in population genetics (in particular, for approximately $s \leq 0.1 $). The denominator can be analogously approximated:
\begin{equation}
\begin{split}
    g_i&=\prod_{j=1}^{i}\bigg(1-\tfrac{B_1}{2N}+B_2\tfrac{2j-N}{2N^2}\bigg)=\exp\bigg(-\tfrac1{2N}\bigg(iB_1+B_2\tfrac{i(N-i-1)}{N}\bigg)\bigg)+\mathcal{O}(1/N^2)\,.
\end{split}
\end{equation}
Therefore the approximate drift and selection terms are given by:
\begin{equation}
    \frac{f_i}{g_i}\approx\frac{\exp\bigg(\tfrac1{2N}\bigg(iB_1+B_2\tfrac{i(N-i+1)}{N}\bigg)\bigg)}{\exp\bigg(-\tfrac1{2N}\bigg(iB_1+B_2\tfrac{i(N-i-1)}{N}\bigg)\bigg)}=\exp\bigg(B_1\tfrac{i}{N}+B_2\tfrac{i(N-i)}{N^2}\bigg)\,.
\end{equation}

\subsubsection{Approximate exponential transition rates and stationary distribution}
\label{app:exponential_stdistr}

We can define an approximate interior transition rates for the boundary-mutation Moran model with linear and quadratic selection, and biased mutation as follows:
\begin{equation}\label{eq:transition_decoupled_Moran_quad_sel}
\begin{cases}
    \Pr(x(t+1)=\frac{i-1}N\given x(t)=\frac{i}{N})&=e^{-\tfrac{B_1}{2N}+B_2\tfrac{2i-N}{2N^2}}\frac{i(N-i)}{N^2}\\
    \Pr(x(t+1)=\frac{i}{N}\given x(t)=\frac{i}{N})&=1-(e^{-\tfrac{B_1}{2N}+B_2\tfrac{2i-N}{2N^2}}+e^{\tfrac{B_1}{2N}-B_2\tfrac{2i-N}{2N^2}})\frac{i(N-i)}{N^2}\\
    \Pr(x(t+1)=\frac{i+1}N\given y(t)=\frac{i}{N})&=e^{\tfrac{B_1}{2N}-\tfrac{B_2(2i-N)}{2N^2}}\frac{i(N-i)}{N^2}\,.
\end{cases}
\end{equation}

The boundary at $i=0$ becomes:
\begin{equation}\label{eq:transition_decoupled_Moran_quad_sel_0}
\begin{cases}
    \Pr(x(t+1)=0\given x(t)=0)&=1-\beta\tfrac{\theta}{N}e^{\tfrac{B_1+B_2}{2N}}D_0\\
    \Pr(x(t+1)=\frac{1}{N}\given x(t)=0)&=\beta\tfrac{\theta}{N}e^{\tfrac{B_1+B_2}{2N}}D_0\,,
\end{cases}
\end{equation}
with 
\begin{equation}
   D_0=\bigg(1-\beta\theta \sum_{i=1}^{N-1} \frac{1}{i} F_i\bigg)^{-1}\,,
\end{equation}
where
\begin{equation}
F_i=e^{B_1\tfrac{i}N+B_2\tfrac{i(N-i)}{N^2}}\,.
\end{equation}
Analogously, at the boundary $i=N$ the approximation yields:
\begin{equation}\label{eq:transition_decoupled_Moran_quad_sel_N}
\begin{cases}
    \Pr(x(t+1)=1\given x(t)=1)&=1-(1-\beta)\tfrac{\theta}{N}e^{\tfrac{-B_1+B_2}{2N}}D_1\\
    \Pr(x(t+1)=\frac{N-1}N\given x(t)=1)&=(1-\beta)\tfrac{\theta}{N}e^{\tfrac{-B_1+B_2}{2N}}D_1\,,
\end{cases}
\end{equation}
with
\begin{equation}
    D_1=\bigg(1-(1-\beta)\theta e^{-B_1} \sum_{i=1}^{N-1} \frac{1}{N-i} F_i\bigg)^{-1}\,.
\end{equation}

Set $\varpi=\frac{\beta(1-\beta)\theta }{(1-\beta)+\beta e^{B_1}}$ and $\varrho=\frac{\beta e^{B_1}}{(1-\beta)+\beta e^{B_1}}$. The approximate equilibrium distribution becomes:
\begin{equation}
\label{eq:equilibrium_quadsel}
   \mathbf{\pi}=\Pr(\mathbf{X} = \frac{i}N \given B_1,B_2,\varrho,\varpi)=
    \begin{cases}
    \displaystyle
    (1-\varrho)-\varpi\sum_{i=1}^{N-1} \frac{1}{i} F_i & i=0;\\
   \displaystyle
   \varpi F_i \frac{N}{i(N-i)} & 1\leq i \leq N-1;\\
    \displaystyle \varrho-\varpi\sum_{i=1}^{N-1} \frac{1}{N-i} F_i & i=N.\,.
\end{cases}
\end{equation}

Detailed balance can be shown analogously as in the exact version and the proof is therefore omitted here.

For small $\theta$, this approximate equilibrium of the  boundary-mutation Moran model is an excellent approximation of Wright's equilibrium distribution with general mutation rates \citep{Wrig31} (Fig.~\ref{fig:ContinuousDiscrete}). 

\subsection{Dynamics of the stationary distribution}

\begin{figure}[!ht]
    \begin{center}
    \includegraphics[scale=0.45]{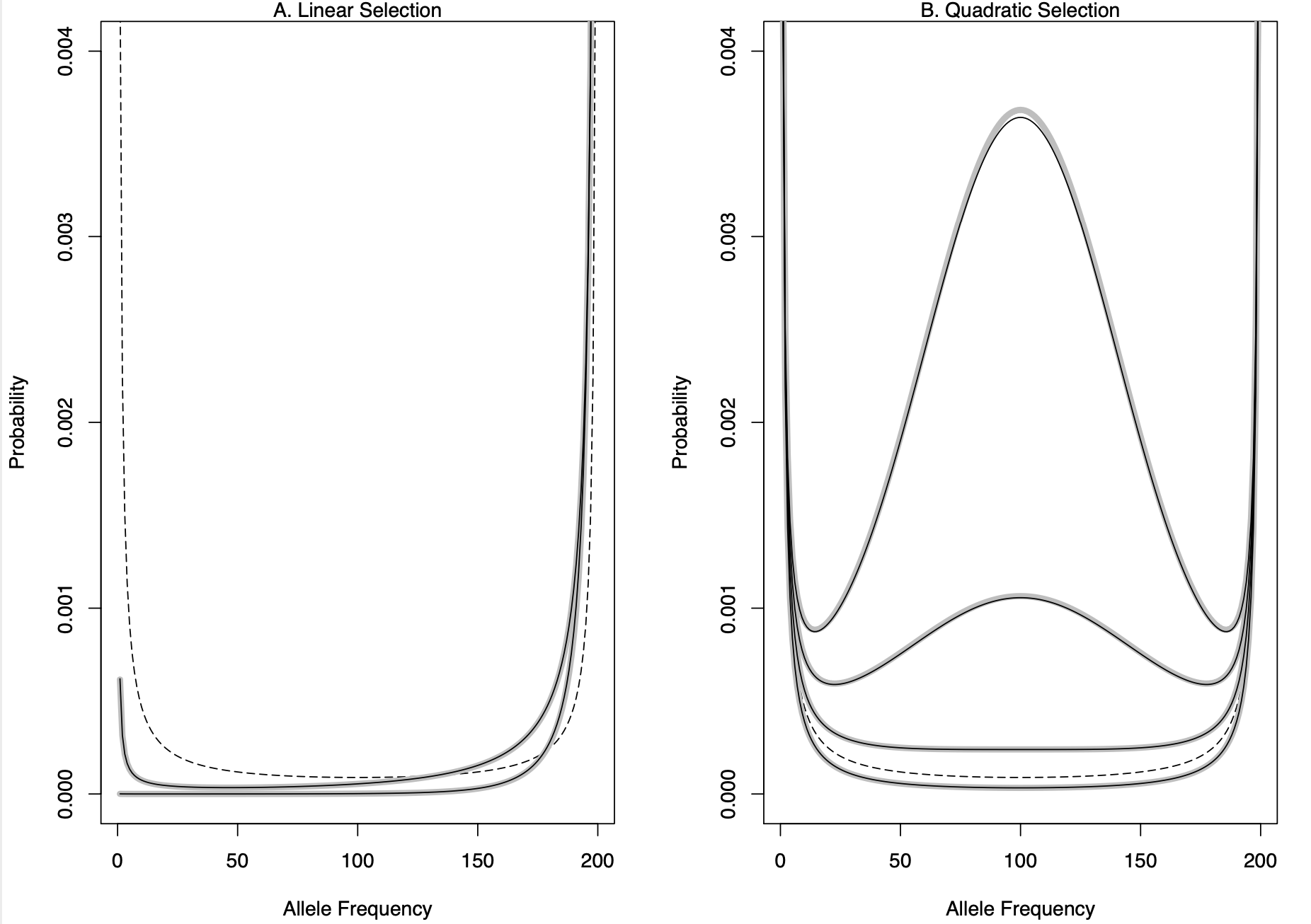}
    \end{center}
    \begin{center}
    \caption{(A.) The equilibrium probability (y-axis) of observing the focal allele within the polymorphic region at a certain frequency (x-axis) under different strengths of linear selection. The neutral scenario is given by the dashed line; the thinner black lines are the exact equilibrium probability for the selection scenarios, the exponential approximation is depicted by the thicker grey lines. Parameters are: $\beta=1/3$, $\theta=0.02$, $N=200$, $B_1=(3,10)$, and $B_2=0$.
    (B.) The equilibrium probability (y-axis) of observing  the focal allele within the polymorphic region at a certain frequency (x-axis) under different strengths of quadratic selection. The neutral scenario is given by the dashed line; the thinner black lines are the exact equilibrium probability for the selection scenarios, the exponential approximation is depicted by the thicker grey lines. Parameters are: $\beta=1/3$, $\theta=0.02$, $N=200$, $B_1=0$, and $B_2=(-4,4,10,15)$.}
    \label{fig:SelectionCoefsAlone}
    \end{center}
\end{figure}

\begin{figure}[!ht]
    \begin{center}
    \includegraphics[scale=0.45]{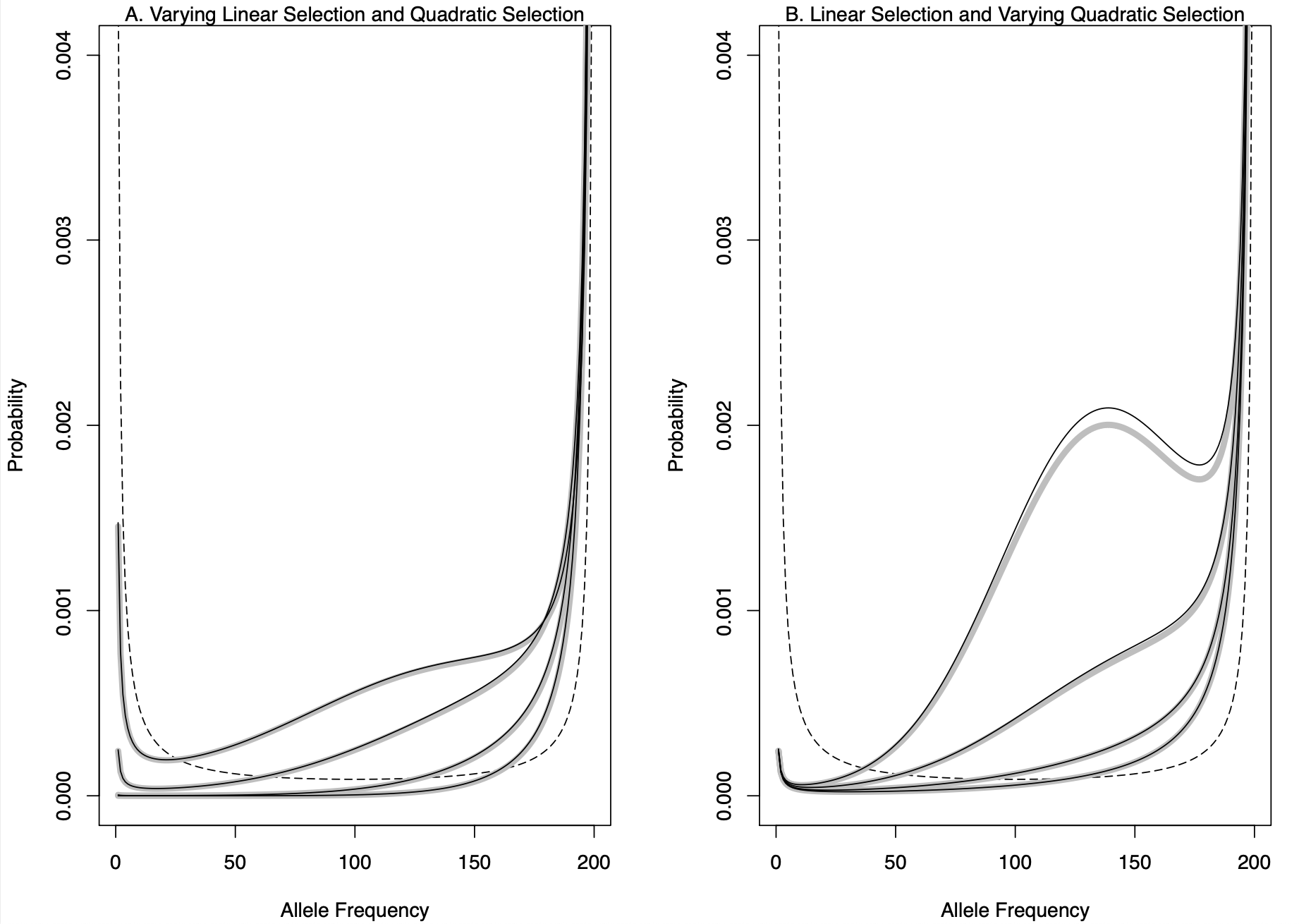}
    \end{center}
    \begin{center}
    \caption{(A.) The equilibrium probability (y-axis) of observing the focal allele within the polymorphic region at a certain frequency (x-axis) under different strengths of linear selection. The neutral scenario is given by the dashed line; the thinner black lines are the exact equilibrium probability for the selection scenarios, the exponential approximation is depicted by the thicker grey lines. Parameters are: $\beta=1/3$, $\theta=0.02$, $N=200$, $B_1=(2,4,8,12)$, and $B_2=8$.
    (B.) The equilibrium probability (y-axis) of observing the focal allele within the polymorphic region at a certain frequency (x-axis) under different strengths of quadratic selection. The neutral scenario is given by the dashed line; the thinner black lines are the exact equilibrium probability for the selection scenarios, the exponential approximation is depicted by the thicker grey lines. Parameters are: $\beta=1/3$, $\theta=0.02$, $N=200$, $B_1=4$, and $B_2=(2,5,10,15)$. }
    \label{fig:SelectionCoefsTogether}
    \end{center}
\end{figure}

Varying the values of $B_1$ and $B_2$, either individually or simultaneously, accounts for a wide range of possible selection scenarios. The exponential distributions are generally good approximations for the exact versions even for very strong selection and small population sizes (Figs.~\ref{fig:SelectionCoefsAlone} and \ref{fig:SelectionCoefsTogether}). 

Note that quadratic selection acts symmetrically around a maximum at frequency $\frac{1}{2}$ when $B_1=0$. Adjusting the latter shifts the target frequency (Figs.~\ref{fig:SelectionCoefsAlone}B and \ref{fig:SelectionCoefsTogether}).
With stabilizing selection around a given optimum, selection may be either mainly directional (far away from the optimum) or underdominant (right at the optimum) \citep{Kimu81}. In the case of overdominance, {\ie} concave fitness on the locus level, a fitness maximum inside the polymorphic region may lead to an increase in polymorphism (Fig.~\ref{fig:SelectionCoefsAlone}B). But in this case the main assumption of the boundary-mutation model, {\ie} that mutations only occur in monomorphic states, may be violated. Recall that the fitness advantage (or disadvantage) through fixation of a mutant allele of the focal type is $B_1$ (irrespective of $B_2$). With the focal allele completely dominant and favored by selection, we have $B_2=B_1$; without dominance $B_2=0$. Hence, dominance makes no difference to zeroth order in $\theta$, {\ie} when drift is strong relative to mutation. This changes when first order terms are included as polymorphism may increase with overdominance and decrease with underdominance, even with relatively low selection coefficients.

\newpage

\newpage

\newpage

\section{Divergence, substitution rates, and heterozygosity}
\label{section:statistics}

In this section formulae for variation within and between populations are derived. We will often use the exponential approximation to the equilibrium distribution from Eq.~(\ref{eq:equilibrium_quadsel}) as a starting point to contrast with neutral versions derived from Eq.~(\ref{eq:equilibrium_nosel}). This is partly for convenience, but we also wish to compare our results to classic diffusion derivations. 

\subsection{Equilibrium distribution among populations}

Let $\varpi\to 0$ in Eq.~(\ref{eq:equilibrium_quadsel}) and compare to the version without selection by letting $\theta\to 0$ in Eq.~(\ref{eq:equilibrium_nosel}). We see from the boundary terms at $i=0$ and $i=N$ of Eq.~(\ref{eq:equilibrium_quadsel}) that the selective advantage of the preferred allele in the entire population is $\gamma=B_1$ to zeroth order in $\theta$. Given $K$ sites with equal effects on the phenotype, a tridiagonal transition rate matrix results and hence we again have detailed balance between nearest neighbours, this time not between alleles within a site but between loci fixed for alternative alleles. Set the number of sites fixed for the focal allele to $y$. In equilibrium the following detailed balance equation must then hold:
\begin{equation}
    \begin{split}
        \Pr(y\given K,\beta,\gamma)(K-y)\beta&=
            \Pr(y+1\given K,\beta,\gamma)(y+1)(1-\beta) e^{-\gamma}\,.     
    \end{split}
\end{equation}

One then sees that the binomial distribution
\begin{equation}\label{eq:binom}
    \Pr(y\given K,\rho)= \binom{K}{y}\rho^y(1-\rho)^{K-y}\,,
\end{equation}
with $\rho=\beta e^{\gamma}/((1-\beta)+\beta e^{\gamma})$ is the equilibrium distribution since
\begin{equation}
    \begin{split}
        \binom{K}{y}\rho^y(1-\rho)^{K-y}(K-y)\beta&=
            \binom{K}{y+1}\rho^{y+1}(1-\rho)^{K-y-1}(y+1)(1-\beta) e^{-\gamma}\\
        (1-\beta)^{K-y}\beta^{y+1}e^{y\gamma}&=
            (1-\beta)^{K-y}\beta^{y+1}e^{(y+1)\gamma}e^{-\gamma}\\
        (1-\beta)^{K-y}\beta^{y+1}e^{y\gamma}&=
            (1-\beta)^{K-y}\beta^{y+1}e^{y\gamma}\,.
    \end{split}
\end{equation}
The mean of the binomial distribution is $K\rho$, the variance among populations $K\rho(1-\rho)$. The variance within populations is zero since each population is assumed fixed at all sites with the first order approximation we made at the start of this subsection. When selection opposes mutation bias, it may increase the variance compared to neutral equilibrium.

Note that $\rho$ corresponds to the expected proportion of favored alleles fixed among the $K$ sites. The equilibrium rates of favored and disfavored new mutations are:
\begin{equation}
\begin{split}
    r_{\beta,\theta,\gamma}&=\sum_{k=0}^K  \frac{K!}{k!(K-y)!}\rho^y(1-\rho)^{K-y} \beta\theta(K-y)=K(1-\rho)\beta\theta\\
    r_{\beta,\theta,-\gamma}&=K\rho(1-\beta)\theta\,.
\end{split}
\end{equation}
The equilibrium ratio of favorable to unfavorable new mutations is independent of the mutation parameters and depends only on selection, as previously noted by \citet{McVe99}:
\begin{equation}\label{eq:ratio_new_muts}
    \frac{r_{\beta,\theta,\gamma}}{r_{\beta,\theta,-\gamma}}=\frac{K(1-\rho)\beta\theta}{K\rho(1-\beta)\theta}=\frac{\beta(1-\beta)\theta}{\beta e^\gamma(1-\beta)\theta}=e^{-\gamma}\,.
\end{equation}
Note that the ratio of the probability of fixation of favorable and unfavorable mutations in equilibrium is $1$. 


\subsection{Expected Heterozygosity}
\label{section:het}

Starting from the boundary-mutation Moran model, an expression for the expected level of heterozygosity can easily be determined. We point out connections between this result, Kimura's formula for heterozygosity \citep{Kimu69a}, and the Ewens-Watterson estimator for molecular diversity \citep{Ewen72,Ewen74,Watt75}.  

\subsubsection{Neutral expected heterozygosity}

In the context of the boundary-mutation Moran model, we can simply sum over the polymorphic region of the equilibrium distribution to obtain a formula for the expected heterozygosity (in contrast to the general model). In the past this has been done for the neutral case from Eq.~(\ref{eq:equilibrium_nosel}) \citep{Vogl12}: 
\begin{equation}
    H_{\beta,\theta}=\sum_{i=1}^{N-1} \beta(1-\beta)\theta\frac{N}{i(N-i)}=2\beta(1-\beta)\theta H_{N-1}\,,
\end{equation}
where $H_{N-1}$ is again the harmonic number. 

Note that this result multiplied by the number of loci is essentially a version of the Ewens-Watterson estimator of molecular diversity \citep{Ewen72,Ewen74,Watt75} for biased mutation. The standard derivations of the Ewens-Watterson estimator use forward diffusion on infinite alleles/sites or coalescent arguments, but in Appendix~(\ref{app:Expect_Het_Kimura}) and Appendix~(\ref{app:Expect_Het_Ewens}) we show that the estimator can be easily derived from Kimura's earlier backward diffusion approach as well \citep{Kimu69a}. 

The boundary-mutation Moran model naturally separates monomorphic and polymorphic dynamics. We can approximate the summation over polymorphic sites with an integral by replacing the allele frequency with the allele proportion $y=i/N$ and taking the limit $N\to\infty$: 
\begin{equation}
    H_{\beta,\theta}=\lim\limits_{N\to\infty} \int_{1/N}^{1-1/N} 2x(1-x)\beta(1-\beta)\theta\frac{1}{x(1-x)}\,dx=2\beta(1-\beta)\theta\,.
\end{equation}
This is then identical to the neutral measure of heterozygosity determined via the Komolgorov backward diffusion  \citep{Kimu69a}.

\subsubsection{Expected heterozygosity under linear selection}

The first derivation of expected heterozygosity under linear selection is due to \citet{Kimu69a}. He assumed the infinite sites model and used boundary-mutation reasoning: A single mutant allele initially segregates at a proportion of $1/N$ (or $1-1/N$). With the Komolgorov backward diffusion equation the course of the allele proportion within the polymorphic region is modeled between $1/N$ and $1-1/N$ (although Kimura integrated from $0$ to $1$ for simplicity), conditional on drift and selection. The reason for neglecting the monomorphic region is that passing a boundary model to the diffusion limit leads to inconsistencies: The diffusion approximation requires the assumption of  $N\to\infty$ for the polymorphic interior. The same assumption causes negative, and thus impossible, probabilities of occupancy at the boun\-da\-ries as discussed in Sec.~\ref{section:historical_intro_pt2}. This can easily be seen from our formulae for equilibrium distributions, e.g. Eq.~(\ref{eq:equilibrium_nosel}), Eq.~(\ref{eq:equilibrium_linsel}), Eq.~(\ref{eq:equilibrium_quadsel}).

Let us now look at the boundary-mutation Moran model with linear selection, {\ie} $\gamma= B_1$, as well as mutation. We will use the exponential approximation of the equilibrium distribution (Eq.~(\ref{eq:equilibrium_quadsel})) to derive the expected heterozygosity and immediately  approximate the sum over the polymorphic region by an integral in order to more readily compare to the diffusion approach. The expected heterozygosity for $N\to\infty$ is then \citep{Vogl15}:
\begin{equation}\label{eq:Heterozygosity}
    \begin{split}
        H_{\beta,\theta,\gamma}&=\lim\limits_{N\to\infty} \int_{1/N}^{1-1/N} 2 x(1-x)\frac{\beta(1-\beta) \theta}{(1-\beta)+\beta e^{\gamma}} e^{\gamma x} \frac{1}{x(1-x)}\,dx\\
        &=2\frac{\beta(1-\beta) \theta}{(1-\beta)+\beta e^{\gamma}}\,\frac{e^{\gamma}-1}{\gamma}=2\varpi\,\frac{e^{\gamma}-1}{\gamma}\,,
    \end{split}
\end{equation}
We show this is equivalent to Kimura's result \citep{Kimu69a} in Appendix~(\ref{app:Expect_Het_Kimura}).

\begin{figure}[h!]
    \begin{center}
    \includegraphics[scale=0.5]{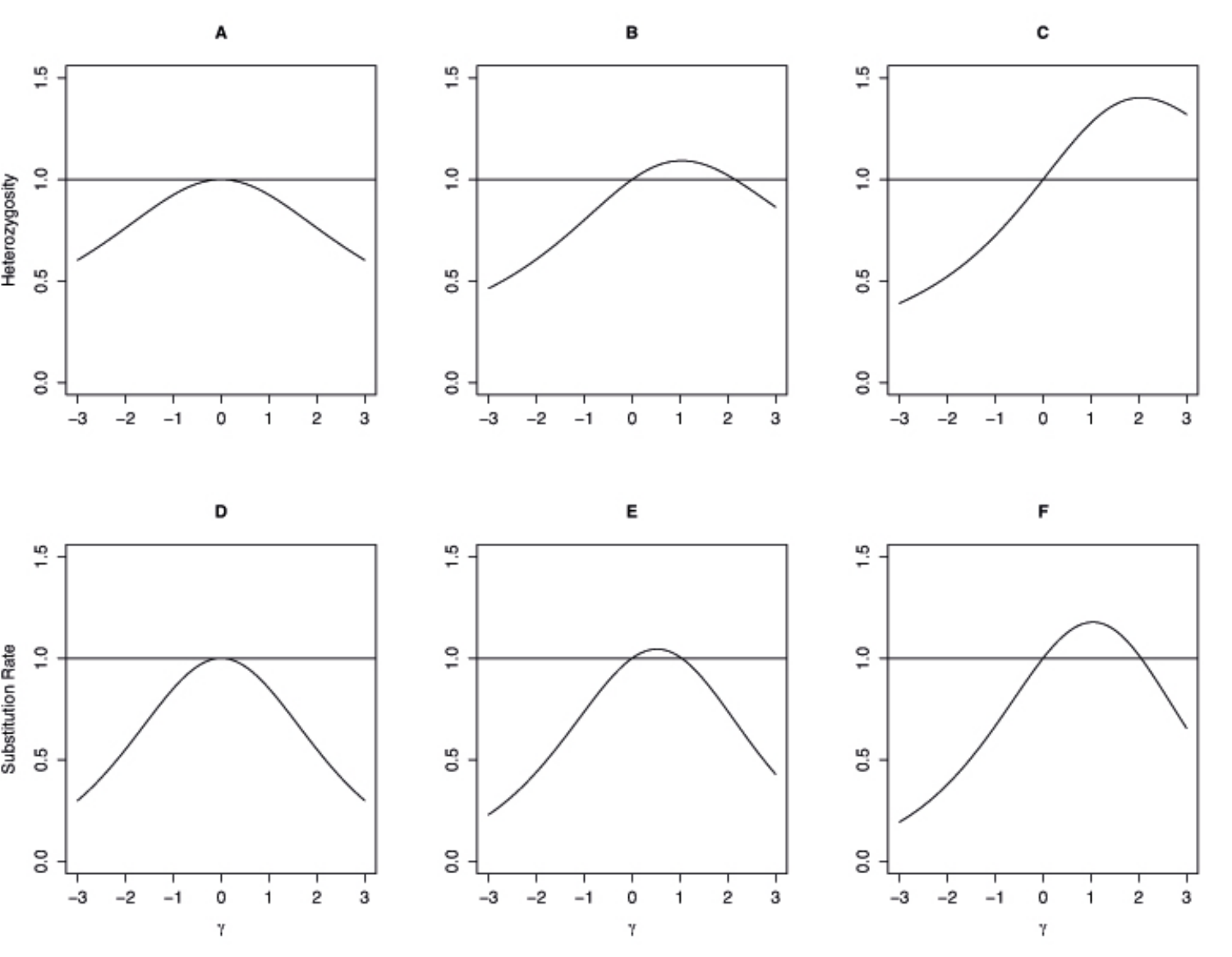}
    \end{center}
    \begin{center}
    \caption{Heterozygosity (A-C) and substitution rate (D-F) relative to neutrality (Eq.~\ref{eq:homozygositiy_linear_selection}, and Eq.~\ref{eq:ratio_of_substitution_rates} respectively), depending on directional selection $\gamma$ for a mutation bias of: $\beta=1/2$ for A and D, $\beta=1/3$ for B and E, $\beta=1/5$ for C and F.}
    \label{fig:SubstitutionRate}
    \end{center}
\end{figure}

In order to detect the action of putative adaptive evolution, the ratio of the expected heterozygosity under linear selection to the expected heterozygosity at neutrality must be evaluated. Using the diffusion approximation, this is given by:
\begin{equation}\label{eq:homozygositiy_linear_selection}
    \frac{H_{\beta,\theta,\gamma}}{H_{\beta,\theta}}=\frac{2\frac{\beta(1-\beta) \theta}{(1-\beta)+\beta e^{\gamma}}\,\frac{e^{\gamma}-1}{\gamma}}{2\beta(1-\beta)\theta}=\frac{1}{(1-\beta)+\beta e^{\gamma}}\,\frac{e^{\gamma}-1}{\gamma}\,.
\end{equation}
While directional selection always decreases heterozygosity when mutation rates are unbiased, directional selection opposing mutation bias may increase heterozygosity (Fig.~\ref{fig:SubstitutionRate}A-C). This happens because directional selection increases the overall mutation rate by favoring the allele with the higher mutation rate. Note that Eq.~(\ref{eq:homozygositiy_linear_selection}) is identical to that given by \citet{McVe99} (see also Appendix~(\ref{app:Expect_Het_McVean_Charlesworth}) for a comparison).


\subsubsection{Expected heterozygosity under quadratic selection}

In continuous models, derivations of heterozygosity that include over- and underdominance involve solving the Gaussian error function. Within the framework of the boundary-mutation Moran model, a sum is taken over the polymorphic region of the approximate equilibrium distribution for the boundary-mutation Moran model with quadratic selection, instead of an integral (Eq.~(\ref{eq:equilibrium_quadsel})): 

Set $\lambda= B_1+\frac{B_2}{N}$ and recall 
$\varpi=\frac{\beta(1-\beta)\theta }{(1-\beta)+\beta e^{B_1+\frac{B_2}{N}}}$. Then:
\begin{equation}
    H_{\beta,\theta,\lambda}=\sum_{i=1}^{N-1} \varpi e^{B_1\tfrac{i}N+B_2\tfrac{i(N-i)}{N^2}}\frac{N}{i(N-i)}= \varpi A_{N-1}\,,
\end{equation}
where $A_{N-1}=\sum_{i=1}^{N-1} e^{B_1\tfrac{i}N+B_2\tfrac{i(N-i)}{N^2}}\frac{N}{i(N-i)}$. The expected heterozygosity under linear and quadratic selection relative to neutrality then becomes:
\begin{equation}\label{eq:homozygositiy_quad_selection}
    \frac{H_{\beta,\theta,\lambda}}{H_{\beta,\theta}}=\frac{1}{2((1-\beta)+\beta e^{\gamma})}\frac{A_{N-1}}{H_{N-1}} \,.
\end{equation}

\begin{figure}[h!]
    \begin{center}
    \includegraphics[scale=0.5]{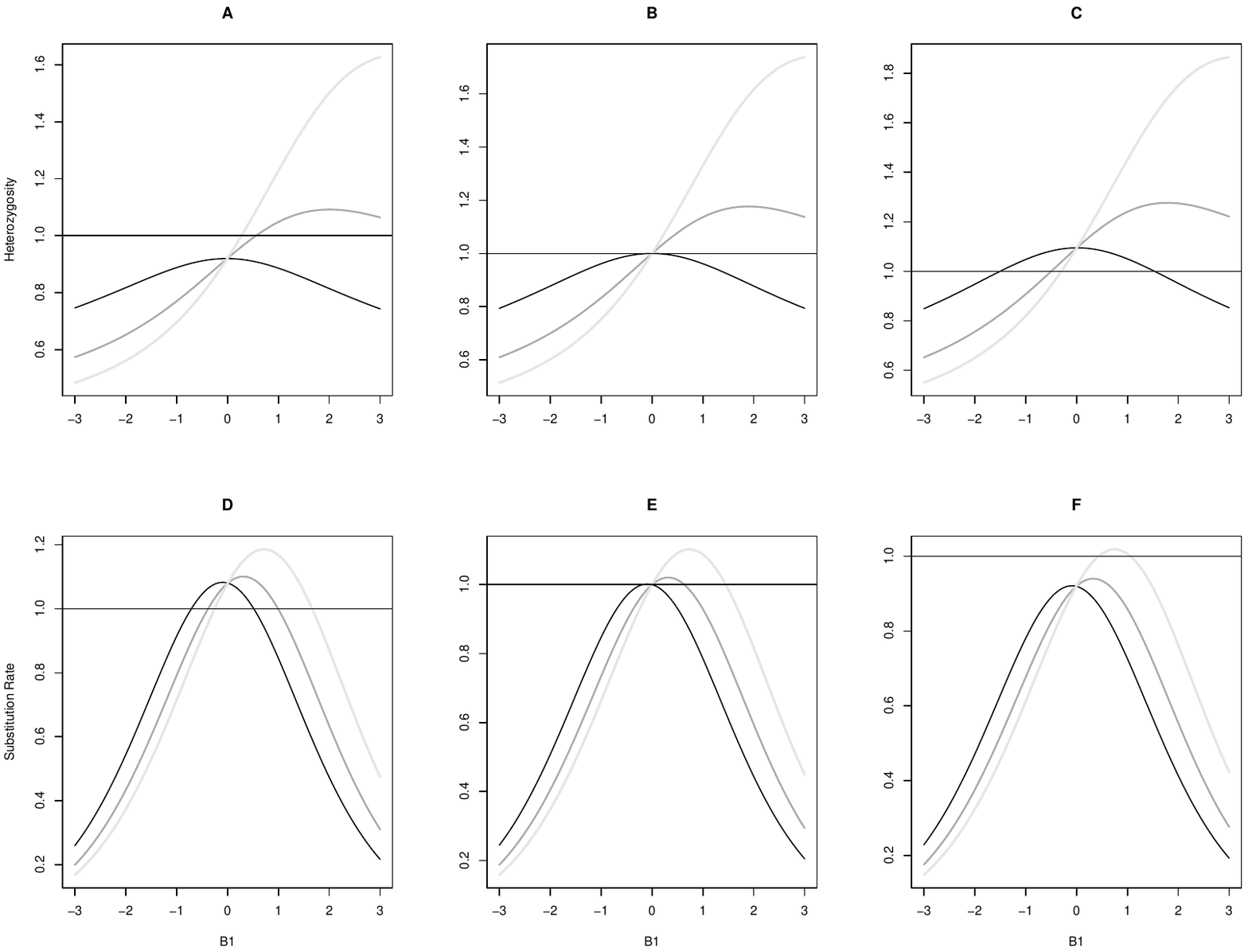}
    \end{center}
    \begin{center}
    \caption{Heterozygosity (A-C) and substitution rate (D-F) relative to neutrality (Eq.~\ref{eq:homozygositiy_quad_selection}, and Eq.~\ref{eq:ratio_of_substitution_rates_quad} respectively) depending on directional selection $B_1$ for $B_2=-1$ for A and D, $B_2=0$ for B and E, $B_2=1$ for C and F. The black line corresponds to $\beta=1/2$, the grey line to $\beta=1/3$, and the light grey line to $\beta=1/5$.}
    \label{fig:SubstitutionRate2}
    \end{center}
\end{figure}

\begin{figure}[h!]
    \begin{center}
    \includegraphics[scale=0.5]{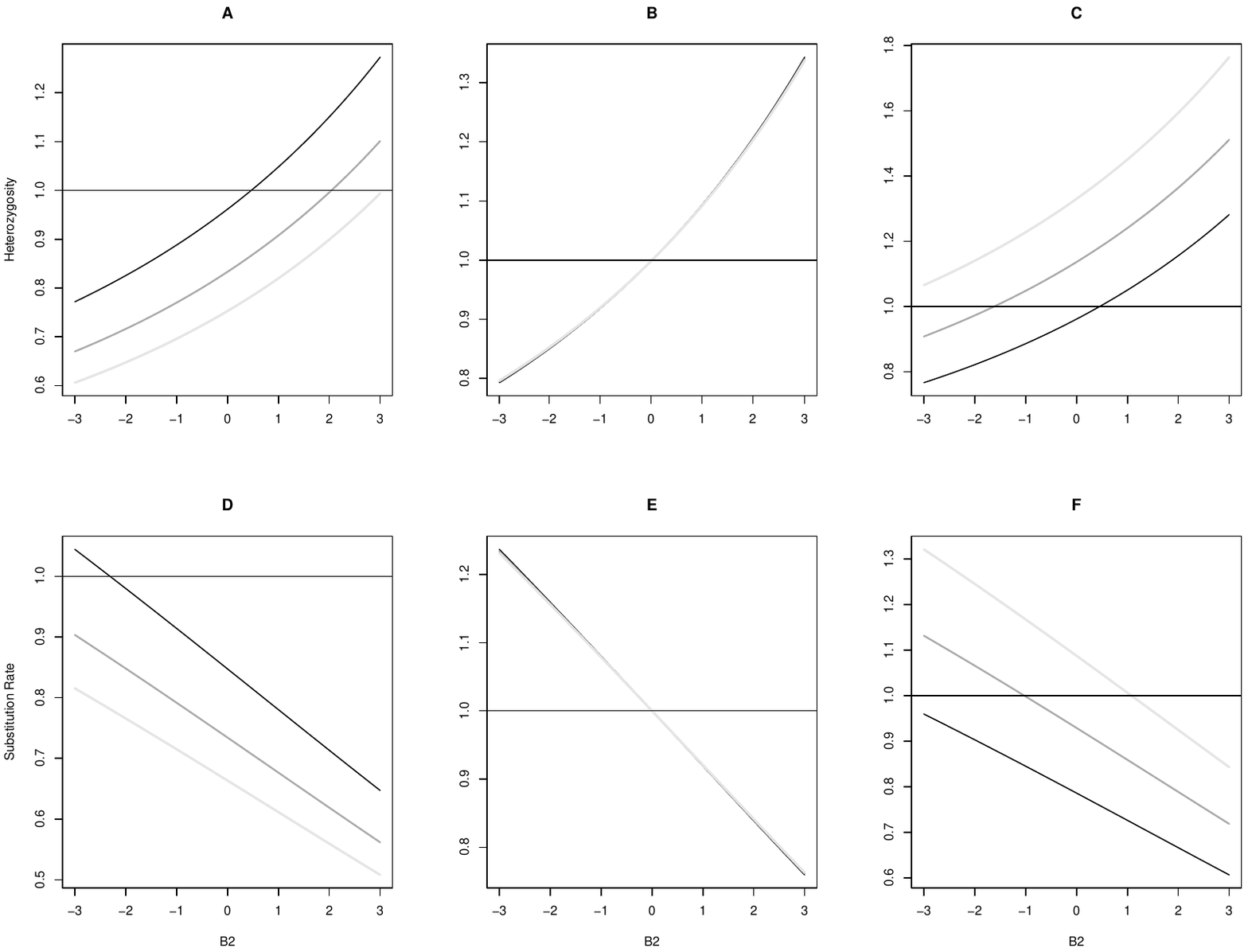}
    \end{center}
    \begin{center}
    \caption{Heterozygosity (A-C) and substitution rate (D-F) relative to neutrality (Eq.~\ref{eq:homozygositiy_quad_selection}, and Eq.~\ref{eq:ratio_of_substitution_rates_quad} respectively) depending on directional selection $B_2$ for $B_1=-1$ for A and D, $B_1=0$ for B and E, $B_1=1$ for C and F. The black line corresponds to $\beta=1/2$, the grey line to $\beta=1/3$, and the light grey line to $\beta=1/5$.}
    \label{fig:SubstitutionRate3}
    \end{center}
\end{figure}

In Fig.~(\ref{fig:SubstitutionRate2}A-C), we see that for a fixed value of quadratic selection the dynamics between linear selection and mutation bias remain the same as without quadratic selection. Relative to neutrality, there is a shift towards lower heterozygosity with negative $B_2$ and towards higher heterozygosity with positive $B_2$ . Fig.~(\ref{fig:SubstitutionRate3}B) makes it apparent that mutation bias only affects quadratic selection if it acts jointly with linear selection. For a fixed value of linear selection, quadratic selection will increase heterozygosity convexly with increasing strength (see Fig.~\ref{fig:SubstitutionRate3}A,C).


\newpage

\newpage

\newpage

\newpage

\subsection{Substitution rates}
\label{sec:sub_rates}

In this subsection, we examine the dynamics of substitution rates, first for varying strengths of mutation bias and opposing linear selection, and then for quadratic selection.

\subsubsection{Fixation probabilities in the boundary-mutation Moran model}

The fixation probability of a new mutation that initially segregates at the boundary and comes under both linear and quadratic selection is given by {\eg} Eq.~(15) of \citet{Kimu81} as: 
\begin{equation}
    \begin{split}
        u_{\beta,\theta,\lambda}\bigg(x=\frac{1}{N}\bigg) &=\frac{1}{N \int_{0}^1 e^{-B_1 x+B_2 x(1-x)}\,dx}\,.
    \end{split}
\end{equation}
This result is derived using the Kolmogorov backward diffusion. 

We have previously relied on Markov process arguments for the majority of our calculations: In \citep{Vogl12}, the fixation rates for a boundary-mutation Moran model with linear selection are determined via balancing conditional flows. However, this is more cumbersome than solving the following discrete difference equation for the fixation probabilities:
\begin{equation}
    \begin{split}
    \label{fix_Kim}
        u_{\beta,\theta,\lambda}(i) &=
        u_{\beta,\theta,\lambda}(i-1) \Pr \bigg(x(t+1)= \frac{i-1}{N} \given x(t)= \frac{i}{N} \bigg)\\
         + & u_{\beta,\theta,\lambda}(i) \bigg(1-\Pr\bigg(x(t+1)= \frac{i-1}{N}  \given x(t)= \frac{i}{N} \bigg)\\
        &- \Pr\bigg(x(t+1)= \frac{i+1}{N} \given x(t)=\frac{i}{N}\bigg) \bigg)\\
        + &u_{\beta,\theta,\lambda}(i+1) \Pr\bigg(x(t+1)= \frac{i+1}{N} \given x(t)=\frac{i}{N}\bigg)\,,
    \end{split}
\end{equation}
where the transition rates are from the exact equilibrium distribution Eq.~(\ref{eq:equilibrium_linsel}).

This is of course the precise discrete equivalent in method to that of Kimura (see also \citep{Kimu62}). Given the boundary conditions $u_{\beta,\theta,\lambda}(0)=0$ and $u_{\beta,\theta,\lambda}(1)=1$, the general result for $i=1,...,N$ is:
\begin{equation}
    \begin{split}
         u_{\beta,\theta,\lambda}(j)= \frac{1+\sum_{j=1}^{i-1} e^{-B_1 \frac{j}{N}+B_2 \frac{j(N-j)}{N^2}}\frac{N}{i(N-i)}}{1+\sum_{j=1}^{N-1} e^{-B_1 \frac{j}{N}+B_2 \frac{j(N-j)}{N^2}}\frac{N}{i(N-i)}}\,, 
      \end{split}
\end{equation}
and, most importantly, for the case of a single segregating mutation this yields: 
\begin{equation}
    \begin{split}
    \label{fix_exact}
        u_{\beta,\theta,\lambda}\bigg(x=\frac{1}{N}\bigg) = \frac{1}{1+\sum_{j=1}^{N-1} e^{-B_1 \frac{j}{N}+B_2 \frac{j(N-j)}{N^2}}\frac{N}{i(N-i)}}\,.
    \end{split}
\end{equation}

\subsubsection{Substitution rates under linear selection and the neutrality index}

With only linear selection, {\ie} $\gamma=B_1$, the substitution rate per generation in equilibrium is balanced between favorable and deleterious mutations. The mutation rate from allele $0$ to $1$ is $\frac{\beta(1-\beta)\theta}{\beta e^\gamma +(1-\beta)}$. If we approximate the fixation probability in Eq.~(\ref{fix_exact}) with the continuous version equivalent in Eq.~(\ref{fix_Kim}), the substitution rate per generation from allele $0$ to allele $1$ is:
\begin{equation}
    \begin{split}
        \frac{\beta(1-\beta)\theta}{\beta e^\gamma +(1-\beta)} \,u_{\beta,\theta,\gamma}\bigg(x=\frac{1}{N}\bigg) &\approx \frac{\beta(1-\beta)\theta}{\beta e^\gamma +(1-\beta)}\,\frac{1}{N \int_{0}^1 e^{-\gamma x}\,dx}\\
        &= \frac{\beta(1-\beta)\theta}{\beta e^\gamma +(1-\beta)}\,\frac{\gamma}{N(1-e^{-\gamma})}\,.
    \end{split}
\end{equation}
In the reverse direction we have $e^\gamma$ more mutations, but the selection coefficient is reversed:
\begin{equation}
    \begin{split}
        \frac{\beta(1-\beta)\theta e^\gamma}{\beta e^\gamma +(1-\beta)} \,u_{\beta,\theta,-\gamma}\bigg(x=\frac{1}{N}\bigg) &\approx \frac{\beta(1-\beta)\theta e^\gamma}{\beta e^\gamma +(1-\beta)}\,\frac{1}{N \int_{0}^1 e^{\gamma x}\,dx}\\
        &= \frac{\beta(1-\beta)\theta}{\beta e^\gamma +(1-\beta)}\,\frac{\gamma e^\gamma}{N(e^{\gamma}-1)}\\
        &=\frac{\beta(1-\beta)\theta}{\beta e^\gamma +(1-\beta)}\,\frac{\gamma}{N(1-e^{-\gamma})}\,.
    \end{split}
\end{equation}

 Recall that $\rho=\beta e^{\gamma}/((1-\beta)+\beta e^{\gamma})$. The overall substitution rate then becomes:
\begin{equation}
    \begin{split}
        S_{\beta,\theta,\gamma}&= 2\sum_{k=0}^K\frac{K!}{k!(K-k)!}\rho^k(1-\rho)^{K-k}
            \,k(1-\beta)\theta\frac{\gamma/N  }{e^{\gamma}-1}\\
            &=2K\frac{\theta}N \frac{\gamma}{e^{\gamma}-1}\,(1-\beta)\rho\,.
    \end{split}
\end{equation}
Without selection this rate reduces to 
\begin{equation}
    \begin{split}
        S_{\beta,\theta}
            &=2K\frac{\theta}N\beta(1-\beta)\,.
    \end{split}
\end{equation}
The ratio of the nearly-neutral and neutral rates is then:
\begin{equation}\label{eq:ratio_of_substitution_rates}
    \begin{split}
    \frac{S_{\beta,\theta,\gamma}}{S_{\beta,\theta}}&=\frac{\rho}{\beta}\frac{\gamma}{e^\gamma-1}=\frac{\gamma}{(\beta e^\gamma+(1-\beta))(1-e^{-\gamma})}\,.
    \end{split}
\end{equation}
Note that while $\frac{\gamma}{e^\gamma-1}$ is always smaller than $1$ for $\gamma>0$, $\frac{\rho}{\beta}$ may be larger because linear selection opposing the mutation bias increases overall mutation rates. Altogether, linear selection may thus increase substitution rates over the neutral rate with biased mutation (Fig.~\ref{fig:SubstitutionRate}D-F).

Following \citet{Ohta72}, the common understanding of selection against deleterious mutations is that it slows down the substitution rate and thus the rate of divergence in proportion to the effective population size. Usually substitution rates elevated above the neutral rate are interpreted as resulting from recent positive selection and not from an equilibrium of linear selection and biased mutation. Yet we are not the first to note that in equilibrium, a strong mutation bias against the optimal codon may actually increase the substitution rate: \citet{McVe99} revived the investigation into this phenomenon; \citet{Lawrie11} also discuss how this interplay can confound maximum likelihood estimates of branch lengths and therefore inference of positive selection on phylogenies.

In Appendix~(\ref{app:Kimura2us}), we provide a comparison between our Eq.~(\ref{eq:ratio_of_substitution_rates}) and the equivalent formula derived by \citet{Kimu81}.


We note that the neutrality index for linear selection vs.\ neutrality is independent of the mutation parameters: It is $1$ for $\gamma=0$, and always greater than $1$ for $\gamma\neq 0$. 
This can be shown as follows:
\begin{equation}\label{eq:neutrality_index}
\begin{split}
    \frac{H_{\beta,\theta,\gamma}}{H_{\beta,\theta}}\frac{S_{\beta,\theta}}{S_{\beta,\theta,\gamma}}
    &=
    \frac{1}{(1-\beta)+\beta e^{\gamma}}\,\frac{e^{\gamma}-1}{\gamma}\,
    \frac{(\beta e^\gamma+(1-\beta))(1-e^{-\gamma})}{\gamma}\\
    &=\frac{e^{\gamma}-1}{\gamma}\,
    \frac{1-e^{-\gamma}}{\gamma}\\
    &=\bigg(\frac{e^{\gamma/2}-e^{-\gamma/2}}{\gamma}\bigg)^2\\
    &=\bigg(\sum_{i=0}^{\infty}\frac{(\gamma/2)^{2i}}{(2i+1)!}\bigg)^2\,.
\end{split}
\end{equation}

\begin{figure}[h!]
    \begin{center}
    \includegraphics[scale=0.4]{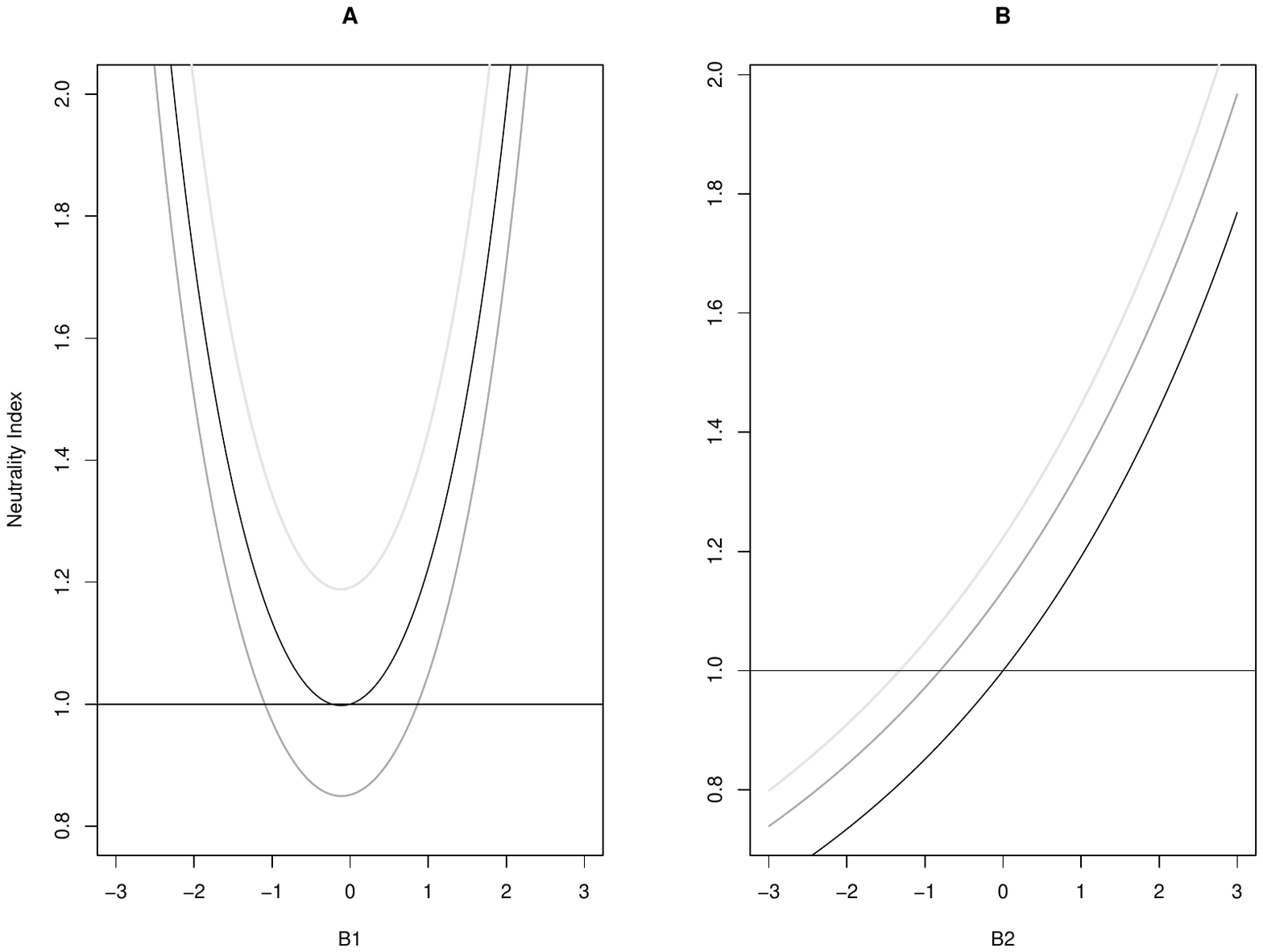}
    \end{center}
    \begin{center}
    \caption{Neutrality index in dependence of $B_1$ (A) and $B_2$ (B) relative to neutrality (Eq.~\ref{eq:neutrality_index_quad}). In (A), the black line corresponds to $B_2=0$, the grey line to $B_2=-1$, and the light grey line to $B_2=1$. In (B), the black line corresponds to $B_1=0$, the grey line to $B_1=-1$, and the light grey line to $B_1=1$. }
    \label{Fig:NI}
    \end{center}
\end{figure}

The squared power series contains only even powers of $\gamma$, such that the neutrality index must be greater than $1$ (since the term for $i=0$ is always $1$). Indeed, the neutrality index is unchanged by reversing the sign of $\gamma$, which actually corresponds to an exchange of the labels of the two alleles (Fig.~\ref{Fig:NI}).


\subsubsection{Substitution rates under quadratic selection and the neutrality index}

As with heterozygosity, obtaining an expression for the substitution rate that includes quadratic as well as linear selection within the framework of the boundary-mutation Moran model involves summation rather than solving the Gaussian error function. We now have $\lambda= B_1+\frac{B_2}{N}$. 
Set $\rho_2=\beta/((1-\beta)+\beta e^{\lambda})$. Then the substitution rate is given by:
\begin{equation}
    \begin{split}
        S_{\beta,\theta,\lambda}&=2\sum_{k=0}^K \frac{K!}{k!(K-k)!}\rho_2^k(1-\rho_2)^{K-k} k(1-\beta)\theta u_{\beta,\theta,\lambda}\bigg(x=\frac{1}{N}\bigg)\\
            &=2K\frac{\theta}N \rho_2 u_{\beta,\theta,\lambda}\bigg(x=\frac{1}{N}\bigg)\,.
    \end{split}
\end{equation}
Without selection this reduces to:
\begin{equation}
    \begin{split}
        S_{\beta,\theta} &=2K\frac{\theta}N\beta(1-\beta)\frac{1}{1+2H_{N-1}}\,.
    \end{split}
\end{equation}
Therefore the substitution rate with quadratic and linear selection relative to neutrality becomes:
\begin{equation}\label{eq:ratio_of_substitution_rates_quad}
    \begin{split}
    \frac{S_{\beta,\theta,\lambda}}{S_{\beta,\theta}}&=\frac{1}{(\beta e^\lambda+(1-\beta))(1+2H_{N-1})u_{\beta,\theta,\lambda}\bigg(x=\frac{1}{N}\bigg)}\,.
    \end{split}
\end{equation}
Inversely as with heterozygosity, quadratic selection causes a shift towards lower substitution rates positive values and towards higher substitution rates for negative ones relative to neutrality (Fig.~\ref{fig:SubstitutionRate2}D-F). This shift appears linear and the slope depends on the interaction between linear selection and mutation bias (Fig.~\ref{fig:SubstitutionRate3}A-C).


Let us take another look at the so-called neutrality index:
\begin{equation}
\begin{split}\label{eq:neutrality_index_quad}
      \frac{H_{\beta,\theta,\lambda}}{H_{\beta,\theta}}\frac{u_{\beta,\theta}}{u_{\beta,\theta,\lambda}}&= \frac{1}{2H_{N-1}} A_{N-1} u_{\beta,\theta,\lambda}\bigg(p=\frac{1}{N}\bigg)\,.
\end{split}
\end{equation}
It is independent of the mutation parameters and has a parabolic shape in dependence on linear selection and a positive curvature in dependence on quadratic selection (Fig.~(\ref{fig:SubstitutionRate2})). Underdominance results in values $<1$ with weak to no linear selection (exact dynamics depending on orientation), whereas overdominance always causes values $>1$.

\newpage

\section{Analysis of Drosophila data}
\label{Sec:Droso_intro}
 
We now apply our formulae to {\em Drosophila} data. This consists of an alignment of ten haploid genomes of Malagasy {\em D.\ simulans} (from inbred isofemale lines) \citep{Rogers14} and ten  haploid genomes of mainland African {\em D.\ melanogaster} (from haploid embryos) \citep{Lack16}, from which we extract a joint site frequency spectrum of fourfold degenerate (FF) and short intronic sites (SI; positions $8-30$bp of introns $<66$bp long) from all autosomal loci. The bases $A$ and $T$ are encoded as allele $0$ and the bases $C$ and $G$ as allele $1$. Then we downsample this spectrum to a sample size of two for {\em D.\ simulans} (columns) and one for {\em D.\ melanogaster} (rows). The joint allele spectrum of short introns is given in Table~(\ref{Table:SI}), where the sum of the second column ({\ie} the sum of the cells [0,1] and [1,1]) corresponds to the heterozygosity and the sum of the cells [0,2] and [1,0] to the divergence. 
\begin{table}[!h]
\caption{Short intron data}
\begin{center}\label{Table:SI}
    \begin{tabular}{l|rrr}
    SI &0 &1 &2\\
    \hline
    0 &56777 &919 & 3145\\
    1 & 2575 &981 &27656\\
    \end{tabular}\\
\end{center}
\end{table}
The joint allele spectrum of the fourfold degenerate sites is given in Table~(\ref{Table:FF}).
\begin{table}[!h]
\caption{Fourfold degenerate site data}
\begin{center}\label{Table:FF}
    \begin{tabular}{l|rrr}
    FF &0 &1 &2\\
    \hline
    0 &291665 & 6616  &35918\\
    1 & 17960 &14848 &613465\\
    \end{tabular}\\
\end{center}
\end{table}

Using only polymorphism and divergence data, a McDonald-Kreitman test of the SI vs.\ FF sites shows a highly significant deviation from neutrality ($\hat\chi^2=44.4$, $p\approx 0$). The FF sites, presumably under selection, are more variable than the presumably neutral SI sites ($\hat\chi^2=10.732$, $p= 0.001$), which is usually interpreted as indicative of balancing selection. As we discussed, as long as the selection direction opposes mutation bias (in this case $\hat \beta_1=0.34$), linear (directional) selection may increase polymorphism.  

In neutral equilibrium, the expected proportions are given in Table~(\ref{Table:exoPropNeutral}), where  $\vartheta =\beta(1-\beta)\theta$, $\tau=\beta(1-\beta)\theta\, t$. Here $t$ is the time of separation of the two species in multiples of $N$ generations.
\begin{table}[h!]
\caption{Expectations under neutrality}
\begin{center}\label{Table:exoPropNeutral}
    \begin{tabular}{l|ccc}
    neutral &0 &1 &2\\
    \hline
    0 &$(1-\beta-\vartheta)(1-\tau)$ &$\vartheta$ &$(\beta-\vartheta)\tau$ \\
    1 &$(1-\beta-\vartheta)\tau$ &$\vartheta$ &$(\beta-\vartheta)(1-\tau)$\\
    \end{tabular}\\
\end{center}
\end{table}
We use the expected proportions in Table~(\ref{Table:exoPropNeutral}) together with the SI data (Table~\ref{Table:SI}) to estimate the parameters: $\hat\vartheta=0.0101$, $\hat\beta=0.345$, and $\hat\tau=0.0625$. From this we conclude that these two species separated $t=\tau/\vartheta=6.17$ generations times the effective population size $N$ ago. 

With selection, the expected proportions are given in Table~(\ref{Table:expPropSel}), with
\begin{equation}
    \eta=\frac{\gamma}{(1-e^{-\gamma})(1-\beta+\beta e^\gamma)}\,,
\end{equation}
\begin{equation}
    \zeta_0=\frac{1}{1-\beta+\beta e^\gamma}\bigg(\bigg(\frac{1}{1-e^{-\gamma}}-\frac{1}{\gamma}\bigg)(1-\vartheta\eta\tau)+\bigg(\frac{e^\gamma}{\gamma}-\frac{1}{1-e^{-\gamma}}\bigg)\vartheta\eta\tau\bigg)\,,
\end{equation} 
and 
\begin{equation}
    \zeta_1= \frac{1}{1-\beta+\beta e^\gamma}\bigg(\bigg(\frac{1}{1-e^{-\gamma}}-\frac{1}{\gamma}\bigg)\vartheta\eta\tau+\bigg(\frac{e^\gamma}{\gamma}-\frac{1}{1-e^{-\gamma}}\bigg) (1-\vartheta\eta\tau)\bigg)\,.
\end{equation}
We note that $\zeta_0$ and $\zeta_1$ are weighted averages of the unidirectional heterozygosity given in Eq.~(\ref{eq:unidirectional_heterozygosity}).
\begin{table}[h!]
\caption{Expectations under selection}
\begin{center}\label{Table:expPropSel}
    \begin{tabular}{l|ccc}
    selected &0 &1 &2\\
    \hline
    0 &$(1-\rho-\vartheta\zeta)(1-\eta\tau)$ &$\vartheta\zeta_0$ &$(\rho-\vartheta\zeta)\eta \tau$ \\
    1 &$(1-\rho-\vartheta\zeta)\eta\tau$ &$\vartheta\zeta_1$ &$(\rho-\vartheta\zeta)(1-\eta\tau)$\\
    \end{tabular}\\
\end{center}
\end{table}
Substituting the estimates $\hat\vartheta$, $\hat\beta$, and $\hat\tau$ into Table~(\ref{Table:expPropSel}) and using the fourfold degenerate site data (Table~\ref{Table:FF}), we obtain an estimator for the directional selection strength of $\hat \gamma=1.364$ by direct search.

\section{Conclusions}
\label{section:Conclusions}

In this article, we introduce a biallelic Moran model with biased, reversible mutation and linear and quadratic selection with mutations from the boundaries only. It approximates the general mutation model when scaled mutation rates are small. We parameterize selection similarly to \citet{Kimu81}, who analyzed a model with many biallelic loci contributing to a normally distributed trait. Our model additionally takes mutation bias into account. While \citet{McVe99} also consider mutation bias, they do not include quadratic selection. In contrast to both, we derive the exact stationary distribution rather than only determining it up to a constant and additionally provide an accurate approximation. This enables particularly direct derivation of expressions for heterozygosity and substitution rates. 

We apply our model to a {\em Drosophila} dataset where sites from short introns are presumably unselected and sites from fourfold degenerate sites are presumably under directional selection. With a McDonald-Kreitman test, heterozygosities and divergence between these two site-classes can be compared and tested for deviation from normality. From the short intron data, we can estimate mutation bias, molecular diversity, and divergence time. Conditional on these estimates, directional selection on fourfold degenerate sites can be determined. We thus go beyond merely demonstrating a deviation from neutrality. We note that molecular diversity is higher for fourfold degenerate sites than for short introns. As long as the direction of directional selection opposes the mutation bias (in this case about $\hat \beta_1=0.34$), linear or directional selection may increase polymorphism. With the inferred parameters, such an effect is expected. It is thus not necessary to postulate balancing selection to explain this result. 

Note that dominance, over- and underdominance, and balancing selection (which are special cases of linear plus quadratic selection) have received a lot of attention in recent years. Many studies have analyzed the signature quadratic selection leaves on linked sites \citep[{\eg}][]{DeGiorgio2014,Bitarello2018}. There are, however, also widely known cases of quadratic selection where linkage does not play a role, such as the human ABO blood polymorphism \citep[][]{Vill15}. 

The reversibility of mutations in our model also implies that the rate of negatively and positively selected new mutations (the ratio of which is determined by mutation bias and selection) may reach a stationary distribution. Thus the effects of deleterious mutations \citep[background selection,][]{CharlesworthMorganCharlesworth93} as well as positively selected mutations \citep[hitchhiking,][]{InnanStephan04} on the effective population size may also equilibrate. Indeed, whether one parameterizes such models with positive or negative selection strength, the resultant equilibrium distribution will be identical. This is a promising setting for future explorations of more complex dynamics involving, {\eg} background selection \citet[see also ideas raised by][]{McVe99}.

In the struggles over the neutral \citep{Kimu83} and nearly-neutral \citep{Ohta96} theories following Gillespie's review \citep{Gill84}, subtleties in the models and common ideas in the approaches of individual scientists took a backseat. Advances were driven by data and computer-intensive approaches. Nowadays, two strands of research appear to prevail within population genetics: one centered on data sets that contrast amino acid changing and silent substitutions \citep{McDo91,Yang00,Smit02}, where theory is based on the infinite sites model with deleterious mutations \citep{Ohta96}; the other centered on data sets that contrast fourfold degenerate sites with short introns \citep{Hadd05,Hadd08,Clem12a,Clem12b,Jack17}, where theory is based on biallelic, reversible mutation models and selection-mutation-drift equilibrium \citep{Li87,Bulm91,McVe99,Vogl15}. Note that both approaches only allow for linear selection; quadratic selection \citep{Kimu81} is ignored. We posit that reversible mutation models such as ours can provide the framework for rigorous analysis of the interplay between mutational bias and both linear and quadratic selection. Broadening the field of application of such reversible models to scenarios in which they are not traditionally used could bring together modelling approaches that seem to have diverged unnecessarily.

\section{Acknowledgments}

We thank Nick Barton, Juraj Bergman, Rui Borges, Reinhard B\"urger, and Joachim Hermisson for helpful discussions, Carolin Kosiol for proofreading and commenting on the manuscript, and Burcin Yildirim for critically reading the manuscript. We also thank the reviewers for their comments, which changed the structure of the article significantly.

CV's research is supported by the Austrian Science Fund (FWF): DK W1225-B20; LCM's by the School of Biology at the University of St.Andrews and has been partially funded through Vienna Science and Technology Fund (WWTF) [MA016-061].

\newpage

\section{Appendix: Comparison to Earlier Results}
\label{app:comparisons}

\subsection{Expected heterozygosity}
\subsubsection{Expected heterozygosity -  Kimura}\label{app:Expect_Het_Kimura}

\citet{Kimu69a} was the first to derive the expected equilibrium heterozygosity with directional selection in an infinite sites model. He did this by using the Komolgorov backward diffusion equation. He assumed time $t$ runs forward through generations. Then, $p$ is the allele proportion $t$ generations earlier with the mean change in proportion over a time interval denoted as $M_{\delta p}$ and the variance as $V_{\delta p}$. The backward equation is then: 
\begin{equation}\label{eq:Kimu_back}
    \frac{\partial\phi(p,x;t)}{\partial t}=\frac12 V_{\delta p}\frac{\partial^2\phi(p,x;t)}{\partial p^2}+M_{\delta p}\frac{\partial\phi(p,x;t)}{\partial p}\,.
\end{equation}
Next, Kimura takes $\nu_m$ to be the number of sites per generation at which a new mutation appears in the population in a Wright-Fisher model. We instead count the per generation mutations from state $0$ to state $1$ within a Moran model - these occur at rate $\frac{\varrho}{2}$. In our case, $\frac{\varrho}{2} \phi(p,x;t) dx$ then represents the contribution of mutants appearing $t$ generations earlier at initial frequency (or rather proportion) $p$ to the present mutant frequencies within the range $x+dx$ (i.e. from $x$ to $x+dx$). Thus, considering all the contributions made by mutations in the past, the expected number of sites at which the mutants presently fall in the frequency range $x+dx$ is:
\begin{equation}
    \phi(p,x)dx=\bigg( \frac{\varrho}{2} \int_{0}^{\infty} \phi(p,x;t)\, dt\bigg)\,dx\,
\end{equation}

Given that the expected heterozygosity under panmixia is $2x(1-x)$, the number of heterozygotes in the population is then: 
\begin{equation}
\begin{split}
    H(p)&= \int_0^{1} 2x(1-x)\phi(p,x)\,  dx\\
    &=\frac{\varrho}{2} \int_0^{1} \bigg(\int_0^{-\infty} \phi(p,x;t)\,  dt\bigg)\,dx\,.
\end{split}
\end{equation}
(Note that the $dx$ appears in the wrong place in the second line of Eq.~(5) in \citet{Kimu69a}.) Then Kimura uses the open interval $0<x<1$ for convenience, but argues that $1/N$ and $1-1/N$ would be more appropriate. He proceeds by multiplying both sides of the backward diffusion in Eq.~(\ref{eq:Kimu_back}) with $2x(1-x)$ and integrating first over $x$ and subsequently over $t$ to obtain
\begin{equation}
    \phi(p,x;0)=\delta(x-p)\,.
\end{equation}
Altogether, he arrives at
\begin{equation}\label{eq:Kimu_(9)}
      0=\frac12 V_{\delta p}\frac{d^2\phi(p,x)}{d p^2}+M_{\delta p}\frac{d\phi(p,x)}{d p}+\varrho p(1-p)\,,
\end{equation}
which he proceeds to solve for different assumptions on $M_{\delta p}$. Kimura starts his discussion of specific scenarios from the general solution of this diffusion equation with boundary conditions $H(0)=H(1)=0$. This is given as:
\begin{equation}\label{eq:H_of_p_Kimu}
    \begin{split}
        H(p)&=(1-u(p))\int_0^p \psi_H(\xi)u(\xi)\,d\xi+ u(p)\int_p^1 \psi_H(\xi)(1-u(\xi))\,d\xi\,,
    \end{split}
\end{equation}
where
\begin{equation}
    \psi_H(\xi)=2\varrho\xi(1-\xi)\int_0^1 G(x)\,dx\frac{1}{V_{\delta \xi}G(\xi)}\,,
\end{equation}
and the ultimate probability of fixation is
\begin{equation}
    u(p)=\frac{\int_0^p G(x)\,dx}{\int_0^1 G(x)\,dx}\,,
\end{equation}
where
\begin{equation}
    G(x)=e^{-2\int_0^x\frac{M_{\delta\xi}}{V_{\delta\xi}}}\,.
\end{equation}

In our Moran model, we have $M_{\delta p}=(B_1-B_2(2p-1))p(1-p)=\gamma p(1-p)$ and $V_{\delta p}=p(1-p)$. We then obtain:
\begin{equation}
    \psi_{H,\gamma,\varrho}(\xi)=2\frac{\varrho}{\gamma}e^{\gamma \xi}( 1 - e^{- \gamma})\,,
\end{equation}
and
\begin{equation}
    u_{\gamma}(p)=\frac{1-e^{- \gamma p}}{1-e^{- \gamma}}\,,
\end{equation}
where
\begin{equation}
    G_{\gamma}(x)=e^{-\gamma x}\,.
\end{equation}

We now wish to show that our formula for the expected heterozygosity can be obtained to order $\frac{1}{N}$ from the general solution of the backward Komolgorov diffusion. We assume mutations occur only at the boundaries in accordance with our mutation model and in this case also with Kimura. Let us consider the initial allele frequency $p=1/N$ at $t=0$. We will eventually take the limit $N\to\infty$. The result is the probability of observing a polymorphic sample of size two (weighted by the mutation rate) in the current generation conditional on a single polymorphism segregating at time $t=0$. 

We begin with Kimura's general formula for heterozygosity (Eq.~(\ref{eq:H_of_p_Kimu})) and substitute our Moran parameters from above: 
\begin{equation}
    \begin{split}
        H_{\gamma,\varrho}(p)&=(1-u_{\gamma})\int_0^p \psi_{H,\gamma,\varrho}(\xi)u(\xi)\,d\xi+ u_{\gamma}(p)\int_p^1 \psi_{H,\gamma,\varrho}(\xi)(1-u_{\gamma}(\xi))\,d\xi\\
         &=\bigg(1 - \frac{1-e^{- \gamma p}}{1-e^{- \gamma }}\bigg)
        \int_0^p 2\frac{\varrho}{\gamma}e^{\gamma \xi}( 1 - e^{- \gamma }) \bigg(\frac{1-e^{-\gamma \xi}}{1-e^{- \gamma }}\bigg)d\xi\\
        &\quad+\bigg(\frac{1-e^{- \gamma p}}{1-e^{- \gamma }}\bigg)
        \int_p^1 2\frac{\varrho}{\gamma}e^{\gamma \xi}( 1 - e^{- \gamma }) \bigg(1-\frac{1-e^{-\gamma \xi}}{1-e^{- \gamma }}\bigg)d\xi\\
        &= 2\frac{\varrho}{\gamma}\bigg(1 - \frac{1-e^{- \gamma p}}{1-e^{- \gamma }}\bigg)
        \int_0^p (e^{\gamma \xi}-1)d\xi\\
        &\quad +2\frac{\varrho}{\gamma}(1-e^{-\gamma \xi}) \int_p^1 e^{\gamma \xi}d\xi + 2\frac{\varrho}{\gamma}\bigg(1 - \frac{1-e^{- \gamma p}}{1-e^{- \gamma }}\bigg) \int_p^1
        (e^{\gamma \xi}-1) d\xi\\
        &= 2\frac{\varrho}{\gamma}\bigg(1 - \frac{1-e^{- \gamma p}}{1-e^{- \gamma}}\bigg)
        \bigg(\frac{e^{\gamma \xi}}{\gamma}\mid_{0}^{p}-\xi\mid_{0}^{p}\bigg)\\
        &\quad +2\frac{\varrho}{\gamma}(1-e^{-\gamma \xi}) \frac{e^{\gamma \xi}}{\gamma}\mid_{p}^{1} + 2\frac{\varrho}{\gamma}\bigg(1 - \frac{1-e^{- \gamma p}}{1-e^{- \gamma }}\bigg)  \bigg(\frac{e^{\gamma \xi}}{\gamma}\mid_{p}^{1}-\xi\mid_{p}^{1}\bigg)\\
    \end{split}
\end{equation}
Now, let us set $p=\frac{1}{N}$ and take the limit $\lim_{N\rightarrow \infty}$ but retain terms of order $\frac{1}{N}$. In doing so, we approximate all exponentials with a Taylor expansion around $0$:
\begin{equation}\label{eq:unidirectional_heterozygosity}
    \begin{split}
\lim_{N\rightarrow \infty} H_{\gamma,\varrho}(p=\frac{1}{N})&= 2\varrho\bigg(1 - \frac{\gamma}{N(1-e^{-\gamma})}\bigg) \cdot  0 \\
&+2\frac{\varrho}{N} \bigg(1 -\frac{e^{\gamma }-1}{\gamma}\bigg)
- 2\frac{\varrho}{N(1-e^{-\gamma })}\bigg(\frac{e^{ \gamma }-1}{\gamma}-1\bigg)+\mathcal{O}\bigg(\frac{1}{N^2}\bigg)\\
=&2\frac{\varrho}{N}\frac{e^{\gamma }-1}{\gamma}-\bigg(2\frac{\varrho}{\gamma}\frac{e^{ \gamma }}{\gamma}- 2\frac{\varrho}{N(1-e^{- \gamma })}\bigg) +\mathcal{O}\bigg(\frac{1}{N^2}\bigg)\\
=&2\frac{\varrho}{N} \bigg(\frac{1}{1-e^{-\gamma}}-\frac{1}{\gamma}\bigg) +\mathcal{O}\bigg(\frac{1}{N^2}\bigg)\,.
 \end{split}
\end{equation}

So far, the result corresponds to the proportion of the heterozygosity in equilibrium that arose through mutations from allele $0$ in the reversible boundary-mutation model. To obtain the heterozygosity for a reversible model, we need to add the proportion that arises through mutations from allele $1$ in equilibrium:
\begin{equation}\label{eq:Het_VM}
\begin{split}    
    H_{\gamma,\varrho}&=2\frac{\varrho}{N}\bigg(\frac{1}{1-e^{-\gamma}}-\frac{1}{\gamma} + e^\gamma\bigg(\frac{1}{1-e^{\gamma}}+\frac{1}{\gamma}\bigg) \bigg)\\
    &=2\frac{\varrho}{N}\,\frac{e^\gamma-1}{\gamma}\,.
\end{split}
\end{equation}
The final result is identical to our Eq.~(\ref{eq:Heterozygosity}).

\subsubsection{Expected heterozygosity - Ewens}\label{app:Expect_Het_Ewens}

\citet{Ewen04}(Eq.~(9.23)) gives the equilibrium distribution of allele proportions in an infinite sites model with scaled mutation rate $\theta$ and selection coefficient $\alpha$ as:
\begin{equation}
    \Pr(\mathbf{X} = i \mid N,\alpha,\theta)=\theta\frac{e^{\alpha(1-x)}-1}{e^{\alpha}-1}\frac{1}{x(1-x)}\,.
\end{equation}
Multiplying this with $2x(1-x)$ and integrating yields the equilibrium heterozygosity:
\begin{equation}\label{eq:Het_Ewens04}
\begin{split}
    H_{\alpha,\theta}&=\int_0^1 \theta\frac{e^{\alpha(1-x)}-1}{e^{\alpha}-1}\frac{1}{x(1-x)} 2x(1-x)\,dx\\
    &=\int_0^1 \theta\frac{e^{\alpha(1-x)}-1}{e^{\alpha}-1}\,dx\\
    &=\frac{2\theta}{e^{\alpha}-1}\int_0^1 e^{\alpha(1-x)}-1\,dx\\
    &=\frac{2\theta}{e^{\alpha}-1} \bigg(\frac{e^{\alpha}-1}{\alpha}-1\bigg)\\
    &=2\theta\bigg(\frac{1}{\alpha}-\frac{1}{e^{\alpha}-1}\bigg)\,.
\end{split}
\end{equation}
This is identical to Kimura's formula if we set $-\alpha=\gamma$ and $\theta=\varrho$.

\subsubsection{Ewens-Watterson Estimator}\label{app:Expect_Het_Ewens_Watt}

Given the equivalence between Kimura's and Ewens's results for the expected heterozygosity in an infinite sites model, it is of interest to check whether the Ewens-Watterson estimator \citep{Ewen72,Ewen74,Watt75} can be obtained with an approach analogous to that of \citet{Kimu69a}. Starting from Eq.~(\ref{eq:Kimu_(9)}), we do not take the expected heterozygosity as $2x(1-x)$ but more generally draw $y$ alleles from a sample of size $M$:
\begin{equation}
      0=\frac12 V_{\delta p}\frac{d^2\phi(p,x)}{d p^2}+M_{\delta p}\frac{d\phi(p,x)}{d p}+\nu \binom{M}{y}p^y(1-p)^{M-y}\,.
\end{equation}

Assume neutrality and $u(\frac{1}{N})=\frac{1}{N}$. Then:
\begin{equation}
\begin{split}
    \lim_{N\to\infty}\psi_{P,\varrho(\xi)}(p=\frac{1}{N})&=N\varrho \binom{M}{y}\xi^y(1-\xi)^{M-y}\frac{\int_0^1 e^0 \,dx}{\xi(1-\xi)}\\
    &=N\varrho\binom{M}{y}\xi^{y-1}(1-\xi)^{M-y-1}\,.
\end{split}
\end{equation}
Substituting into the analog of Eq.~(\ref{eq:H_of_p_Kimu}), we obtain for $1\leq y\leq M$:
\begin{equation}
    \begin{split}
    \Pr(\mathbf{Y} = y \given M,\varrho)&= \varrho \binom{M}{y}\int_0^1 \xi^{y-1}(1-\xi)^{M-y-1}\,dx\\
    &=\varrho \binom{M}{y}\frac{\Gamma(y)\Gamma(M-y)}{\Gamma(M)}=\varrho \frac{M}{y(M-y)}=\varrho (\tfrac{M}{y}+\tfrac{M}{M-y})\,.
    \end{split}
\end{equation}
Summing over $y$ from $1$ to $M-1$, we obtain:
\begin{equation}
    \begin{split}
    \sum_{y=1}^{M-1}\Pr(\mathbf{Y} = y \given M,\varrho)&= \sum_{y=1}^{M-1}\varrho (\tfrac{M}{y}+\tfrac{M}{M-y})\\
    &=2\varrho \sum_{y=1}^{M-1}\tfrac{1}{y}=2\varrho H_{M-1}\,,
    \end{split}
\end{equation}
where $H_{M-1}$ is the harmonic number. Multiplying by the number of sites $L$, we obtain the expected number of polymorphic sites and the Ewens-Watterson estimator of genetic diversity if $\theta=2\varrho$.

\subsubsection{Expected heterozygosity according to McVean and Charlesworth}\label{app:Expect_Het_McVean_Charlesworth}

\citet{McVe99} give the following approximate formula for the expected heterozygosity in their reversible mutation model (in our notation) as:
\begin{equation}\label{eq:Het_MC}
\begin{split}
    H_{\beta,\theta,\gamma (MC)}&=4\beta\theta \frac{\tfrac{(1-\beta)}{\beta} (e^\gamma-1)}{\gamma\big(\tfrac{(1-\beta)}{\beta}+e^\gamma \big)}\\
\end{split}
\end{equation}
One can easily see that this is identical to our Eq.~(\ref{eq:Het_VM}) up to a factor of two that stems from the difference between the Wright-Fisher and the Moran model.

\subsection{Evolutionary Rate}
\label{app:Kimura2us}

\citet{Kimu81} gives the relative evolutionary rate (in terms of mutant substitutions) under directional selection compared with the strictly neutral case in his Eq.~(25):
\begin{equation}\label{eq:Kimu_ratio_subst}
    S_{\beta,\theta,\gamma}/S_{\beta,\theta}=2f_1f_2\log(f_2/f_1)/(f_2-f_1)\,.
\end{equation}
Here, $f_2$ is the equilibrium proportion of the positively selected allele and $f_1$ that of the other. 

Assuming unbiased mutation, Kimura's Eq.~(22) for the ratio of allele proportions gives $f_2/f_1=e^\gamma$. 

Substituting this into the above Eq.~(\ref{eq:Kimu_ratio_subst}), we get
\begin{equation}
\begin{split}
    S_{\beta,\theta,\gamma}/S_{\beta,\theta}&=2\frac{e^\gamma}{(1+e^\gamma)^2}\log(e^\gamma)\frac{1+e^\gamma}{e^\gamma-1}\\
    &=2\frac{\gamma e^\gamma}{(1+e^\gamma)(e^\gamma-1)}\\
    &=2\frac{\gamma}{(1+e^\gamma)(1-e^{-\gamma})}\,.
\end{split}
\end{equation}

This is identical to our Eq.~(\ref{eq:ratio_of_substitution_rates}) without mutation bias, {\ie} $\beta=\tfrac{1}{2}$. 
With mutation bias, a different formula from ours would result. From the numerical example, it can be concluded that Kimura assumed the absence of mutation bias, to which he generally seems to have given little thought. In this case, the substitution rate cannot increase under directional selection compared to neutrality.

\newpage
\section*{References}

\bibliography{coal}

\newcommand{\noopsort}[1]{} \newcommand{\printfirst}[2]{#1}
  \newcommand{\singleletter}[1]{#1} \newcommand{\switchargs}[2]{#2#1}
\begin{thebibliography}{}

\bibitem[Akashi(1994)Akashi]{Akas94}
Akashi, H. (1994).
\newblock Synonymous Codon Usage in Drosophila melanogaster: Natural Selection
  and Translational Accuracy.
\newblock {\em Genetics Society of America\/}, {\bf 136}, 927--935.

\bibitem[Baake and Bialowons(2008)Baake and Bialowons]{Baak08}
Baake, E. and Bialowons, R. (2008).
\newblock Ancestral processes with selection: branching and Moran models.
\newblock {\em Banach center publications\/}, {\bf 80}, 33--52.

\bibitem[Bergman {\em et~al.}(2018)Bergman, Schrempf, Kosiol, and Vogl]{Berg17}
Bergman, J., Schrempf, D., Kosiol, C., and Vogl, C. (2018).
\newblock Inference in Population Genetics Using Forward and Backward, Discrete
  and Continuous Time Processes.
\newblock {\em Journal of Theoretical Biology\/}, {\bf 439}, 166--180.

\bibitem[Bitarello {\em et~al.}(2018)Bitarello, {De Filippo}, Teixeira,
  Schmidt, Kleinert, Meyer, and Andr{\'{e}}s]{Bitarello2018}
Bitarello, B.~D., {De Filippo}, C., Teixeira, J.~C., Schmidt, J.~M., Kleinert,
  P., Meyer, D., and Andr{\'{e}}s, A.~M. (2018).
\newblock {Signatures of long-term balancing selection in human genomes}.
\newblock {\em Genome Biology and Evolution\/}, {\bf 10}, 939--955.

\bibitem[Bulmer(1991)Bulmer]{Bulm91}
Bulmer, M. (1991).
\newblock The selection-mutation-drift theory of synonymous codon usage.
\newblock {\em Genetics\/}, {\bf 129}, 897--907.

\bibitem[Charlesworth {\em et~al.}(1993)Charlesworth, Morgan, and
  Charlesworth]{CharlesworthMorganCharlesworth93}
Charlesworth, B., Morgan, M., and Charlesworth, D. (1993).
\newblock The effect of deleterious mutations on neutral molecular variation.
\newblock {\em Genetics\/}, {\bf 134}, 1289--1303.

\bibitem[Clemente and Vogl(2012a)Clemente and Vogl]{Clem12b}
Clemente, F. and Vogl, C. (2012a).
\newblock Evidence for complex selection on four-fold degenerate sites in {\it
  Drosophila melanogaster}.
\newblock {\em J.\ Evol.\ Biol.}, {\bf 25}(12), 2582--95.

\bibitem[Clemente and Vogl(2012b)Clemente and Vogl]{Clem12a}
Clemente, F. and Vogl, C. (2012b).
\newblock Unconstrained evolution in short introns?---An analysis of
  genome-wide polymorphism and divergence data from {\it Drosophila}.
\newblock {\em J.\ Evol.\ Biol.}, {\bf 25}(10), 1975--90.

\bibitem[De~Maio {\em et~al.}(2013)De~Maio, Schl\"otterer, and Kosiol]{DeMa13}
De~Maio, N., Schl\"otterer, C., and Kosiol, C. (2013).
\newblock Linking great apes genome evolution across time scales using
  polymorphism-aware phylogenetic models.
\newblock {\em Mol.\ Biol.\ Evol.}, {\bf 30}, 2249--2262.

\bibitem[{De Maio} {\em et~al.}(2015){De Maio}, Schrempf, and
  Kosiol]{DeMaio2015}
{De Maio}, N., Schrempf, D., and Kosiol, C. (2015).
\newblock {PoMo: An Allele Frequency-Based Approach for Species Tree
  Estimation}.
\newblock {\em Systematic Biology\/}, {\bf 64}(6), 1018--1031.

\bibitem[DeGiorgio {\em et~al.}(2014)DeGiorgio, Lohmueller, and
  Nielsen]{DeGiorgio2014}
DeGiorgio, M., Lohmueller, K.~E., and Nielsen, R. (2014).
\newblock {A Model-Based Approach for Identifying Signatures of Ancient
  Balancing Selection in Genetic Data}.
\newblock {\em PLoS Genetics\/}, {\bf 10}(8).

\bibitem[Etheridge and Griffiths(2009)Etheridge and Griffiths]{Ethe09}
Etheridge, A. and Griffiths, R. (2009).
\newblock A coalescent dual process in a Moran model with genic selection.
\newblock {\em Theoretical Population Biology\/}, {\bf 75}, 320--330.

\bibitem[Ewens(1972)Ewens]{Ewen72}
Ewens, W. (1972).
\newblock The sampling theory of selectively neutral alleles.
\newblock {\em Theoretical Population Biology\/}, {\bf 3}, 87--112.

\bibitem[Ewens(1974)Ewens]{Ewen74}
Ewens, W. (1974).
\newblock A note on the sampling theory for infinite alleles and infinite sites
  models.
\newblock {\em Theoretical Population Biology\/}, {\bf 6}, 143--148.

\bibitem[Ewens(2004)Ewens]{Ewen04}
Ewens, W. (2004).
\newblock {\em Mathematical Population Genetics\/}.
\newblock Springer, N.Y., 2nd edition.

\bibitem[Gillespie(1984)Gillespie]{Gill84}
Gillespie, J. (1984).
\newblock The Status of the Neutral Theory.
\newblock {\em Science\/}, {\bf 224}, 732--733.

\bibitem[Haddrill {\em et~al.}(2005)Haddrill, Thornton, Charlesworth, and
  Andolfatto]{Hadd05}
Haddrill, P., Thornton, K., Charlesworth, B., and Andolfatto, P. (2005).
\newblock Multilocus patterns of nucleotide variability and the demographic and
  selection history of Drosophila melanogaster population.
\newblock {\em Genome Research\/}, {\bf 15}, 790--799.

\bibitem[Haddrill and Charlesworth(2008)Haddrill and Charlesworth]{Hadd08}
Haddrill, P.~R. and Charlesworth, B. (2008).
\newblock Non-neutral processes drive the nucleotide composition of non-coding
  sequences in Drosophila.
\newblock {\em Biol Lett\/}, {\bf 4}(4), 438--41.

\bibitem[Ingvarsson(2010)Ingvarsson]{Ingv10}
Ingvarsson, P.~K. (2010).
\newblock {Natural Selection on Synonymous and Nonsynonymous Mutations Shapes
  Patterns of Polymorphism in Populus tremula}.
\newblock {\em Molecular Biology and Evolution\/}, {\bf 27}(3), 650--660.

\bibitem[Innan and Stephan(2002)Innan and Stephan]{InnanStephan04}
Innan, H. and Stephan, W. (2002).
\newblock Distinguishing the Hitchhiking and Background Selection Models.
\newblock {\em Genetics\/}, {\bf 165}, 2307–--2312.

\bibitem[Jackson {\em et~al.}(2017)Jackson, Campos, Haddrill, Charlesworth, and
  Zeng]{Jack17}
Jackson, B., Campos, J., Haddrill, P., Charlesworth, B., and Zeng, K. (2017).
\newblock Variation in the Intensity of Selection on Codon Bias over Time
  Causes Contrasting Patterns of Base Composition Evolution in {\it
  Drosophila}.
\newblock {\em Genome Biology and Evolution\/}.

\bibitem[Karlin and Taylor(1975)Karlin and Taylor]{Karl75}
Karlin, S. and Taylor, H. (1975).
\newblock {\em A First Course In Stochastic Processes\/}.
\newblock Academic Press.

\bibitem[Kimura(1962)Kimura]{Kimu62}
Kimura, M. (1962).
\newblock On the probability of fixation of mutant genes in a population.
\newblock {\em Genetics\/}, {\bf 47}, 713--719.

\bibitem[Kimura(1969)Kimura]{Kimu69a}
Kimura, M. (1969).
\newblock The number of heterozygous nucleotide sites maintained in a finite
  population due to steady flux of mutations.
\newblock {\em Genetics\/}, {\bf 61}, 893--903.

\bibitem[Kimura(1981)Kimura]{Kimu81}
Kimura, M. (1981).
\newblock Possibility of extensive neutral evolution under stabilizing
  selection with special reference to nonrandom use of codons.
\newblock {\em Proceedings of the National Academy of Sciences, USA\/}, {\bf
  78}, 5773--5777.

\bibitem[Kimura(1983)Kimura]{Kimu83}
Kimura, M. (1983).
\newblock {\em The Neutral Theory of Molecular Evolution.}
\newblock Cambridge University Press.

\bibitem[Lack {\em et~al.}(2016)Lack, Lange, Tang, Corbett-Detig, and
  Pool]{Lack16}
Lack, J., Lange, J., Tang, A., Corbett-Detig, R., and Pool, J. (2016).
\newblock A Thousand Fly Genomes: An Expanded Drosophila Genome Nexus.
\newblock {\em Molecular Biology and Evolution\/}, {\bf 33}, 3308–3313.

\bibitem[Lawrie {\em et~al.}(2011)Lawrie, Petrov, and PW]{Lawrie11}
Lawrie, D., Petrov, D., and PW, M. (2011).
\newblock Faster than neutral evolution of constrained sequences: the complex
  interplay of mutational biases and weak selection.
\newblock {\em Genome Biology and Evolution\/}, {\bf 3}, 383--–395.

\bibitem[Lawrie {\em et~al.}(2013)Lawrie, Messer, Hershberg, and
  Petrov]{Lawrie20}
Lawrie, D., Messer, P., Hershberg, R., and Petrov, D. (2013).
\newblock Strong purifying selection at synonymous sites in D. melanogaster.
\newblock {\em PloS Genetics\/}, {\bf 9(5)}, e1003527.

\bibitem[Li(1987)Li]{Li87}
Li, W. (1987).
\newblock Models of nearly neutral mutations with particular implications for
  nonrandom usage of synonymous codons.
\newblock {\em Journal of Molecular Evolution\/}, {\bf 24}, 337--345.

\bibitem[Lynch {\em et~al.}(2016)Lynch, Ackerman, Gout, Long, Sung, Thomas, and
  Foster]{Lynch16}
Lynch, M., Ackerman, M., Gout, J., Long, H., Sung, W., Thomas, W., and Foster,
  P. (2016).
\newblock Genetic drift, selection and the evolution of the mutation rate.
\newblock {\em Nature Reviews Genetics\/}, {\bf 17}, 704--714.

\bibitem[Machado {\em et~al.}(2020)Machado, Lawrie, and Petrov]{Machado19}
Machado, H., Lawrie, D., and Petrov, D. (2020).
\newblock Pervasive strong selection at the level of codon usage bias in {\em
  Drosophila melanogaster}.
\newblock {\em Genetics\/}, {\bf 214(2)}, 511--528.

\bibitem[McDonald and Kreitman(1991)McDonald and Kreitman]{McDo91}
McDonald, J. and Kreitman, M. (1991).
\newblock Adaptive protein evolution at the ADH locus in {\em Drosophila}.
\newblock {\em Nature\/}, {\bf 351}, 652–654.

\bibitem[McVean and Charlesworth(1999)McVean and Charlesworth]{McVe99}
McVean, G. and Charlesworth, B. (1999).
\newblock A population genetic model for the evolution of synonymous codon
  usage: patterns and predictions.
\newblock {\em Genet.\ Res.}, {\bf 74}, 145--158.

\bibitem[Moran(1958a)Moran]{Mora58b}
Moran, P. (1958a).
\newblock The effect of selection in a haploid genetic population.
\newblock {\em Proc.\ Camb.\ Phil.\ Soc.}, {\bf 54}, 463--467.

\bibitem[Moran(1958b)Moran]{Mora58a}
Moran, P. (1958b).
\newblock Random processes in genetics.
\newblock {\em Proc.\ Camb.\ Phil.\ Soc.}, {\bf 54}, 60--71.

\bibitem[Moran(1962)Moran]{Mora62}
Moran, P. (1962).
\newblock {\em Statistical processes of evolutionary theory.}
\newblock Clarendon Press, Oxford.

\bibitem[Muirhead and Wakeley(2009)Muirhead and Wakeley]{Muir09}
Muirhead, C. and Wakeley, J. (2009).
\newblock Modeling multi-allelic selection using a Moran model.
\newblock {\em Genetics\/}, {\bf 182}, 1141--1157.

\bibitem[Nielsen and Yang(2003)Nielsen and Yang]{Niel03}
Nielsen, R. and Yang, Z. (2003).
\newblock Estimating the Distribution of Selection Coefficients from
  Phylogenetic Data with Applications to Mitochondrial and Viral DNA.
\newblock {\em Molecular Biology and Evolution\/}, {\bf 20}, 1231–--1239.

\bibitem[Ohta(1972)Ohta]{Ohta72}
Ohta, T. (1972).
\newblock Evolutionary rate of cistrons and DNA divergence.
\newblock {\em Journal of Molecular Evolution\/}, {\bf 1}, 150--157.

\bibitem[Ohta(1979)Ohta]{Ohta73}
Ohta, T. (1979).
\newblock Slightly deleterious mutant substitutions in evolution.
\newblock {\em Nature\/}, {\bf 246}, 96--98.

\bibitem[Ohta and Gillespie(1996)Ohta and Gillespie]{Ohta96}
Ohta, T. and Gillespie, J. (1996).
\newblock Development of neutral and nearly neutral theories.
\newblock {\em Theoretical Population Biology\/}, {\bf 49}, 128--142.

\bibitem[Rogers {\em et~al.}(2014)Rogers, Cridland, Shao, Hu, Andolfatto, and
  Thornton]{Rogers14}
Rogers, R.~L., Cridland, J.~M., Shao, L., Hu, T.~T., Andolfatto, P., and
  Thornton, K.~R. (2014).
\newblock Landscape of standing variation for tandem duplications in Drosophila
  yakuba and Drosophila simulans.
\newblock {\em Molecular Biology and Evolution\/}, {\bf 31}(7), 1750--66.

\bibitem[Rousselle {\em et~al.}(2019)Rousselle, Lavarre, Figuet, Nabholz, and
  N.]{Rou19}
Rousselle, M., Lavarre, A., Figuet, E., Nabholz, B., and N., G. (2019).
\newblock {Influence of Recombination and GC-biased Gene Conversion on the
  Adaptive and Nonadaptive Substitution Rate in Mammals versus Birds}.
\newblock {\em Molecular Biology and Evolution\/}, {\bf 36}(3), 458--471.

\bibitem[Smith and Eyre-Walker(2002)Smith and Eyre-Walker]{Smit02}
Smith, N. and Eyre-Walker, A. (2002).
\newblock Adaptive protein evolution in Drosophila.
\newblock {\em Nature\/}, {\bf 415}, 1022–1024.

\bibitem[Tachida(1991)Tachida]{Tach91}
Tachida, H. (1991).
\newblock A study on a nearly neutral mutation model in finite populations.
\newblock {\em Genetics\/}, {\bf 128}, 420--433.

\bibitem[Villanea {\em et~al.}(2015)Villanea, Safi, and Busch]{Vill15}
Villanea, F., Safi, K., and Busch, J. (2015).
\newblock General Model of Negative Frequency Dependent Selection Explains
  Global Patterns of Human ABO Polymorphism.
\newblock {\em PLoS ONE\/}, {\bf 10(5)}.

\bibitem[Vogl(2014)Vogl]{Vogl14b}
Vogl, C. (2014).
\newblock Estimating the Scaled Mutation Rate and Mutation Bias with Site
  Frequency Data.
\newblock {\em Theoretical Population Biology\/}, {\bf 98}, 19---27.

\bibitem[Vogl and Bergman(2015)Vogl and Bergman]{Vogl15}
Vogl, C. and Bergman, J. (2015).
\newblock Inference of directional selection and mutation parameters assuming
  equilibrium.
\newblock {\em Theoretical Population Biology\/}, {\bf 106}, 71--82.

\bibitem[Vogl and Clemente(2012)Vogl and Clemente]{Vogl12}
Vogl, C. and Clemente, F. (2012).
\newblock The allele-frequency spectrum in a decoupled Moran model with
  mutation, drift, and directional selection, assuming small mutation rates.
\newblock {\em Theoretical Population Genetics\/}, {\bf 81}, 197--209.

\bibitem[Watterson(1975)Watterson]{Watt75}
Watterson, G. (1975).
\newblock On the number of segregating sites in genetical models without
  recombination.
\newblock {\em Theoretical Population Biology\/}, {\bf 7}, 256--276.

\bibitem[Wright(1931)Wright]{Wrig31}
Wright, S. (1931).
\newblock Evolution in Mendelian populations.
\newblock {\em Genetics\/}, {\bf 16}, 97--159.

\bibitem[Yang and Bialewski(2000)Yang and Bialewski]{YangBial00}
Yang, Z. and Bialewski, J. (2000).
\newblock Statistical methods for detecting molecular adaptation.
\newblock {\em Trends in Ecology and Evolution\/}, {\bf 15}, 496--503.

\bibitem[Yang and Nielsen(1998)Yang and Nielsen]{Yang98}
Yang, Z. and Nielsen, R. (1998).
\newblock Synonymous and Nonsynonymous Rate Variation in Nuclear Genes of
  Mammals.
\newblock {\em Journal of Molecular Evolution\/}, {\bf 46}, 409--418.

\bibitem[Yang and Nielsen(2000)Yang and Nielsen]{Yang00}
Yang, Z. and Nielsen, R. (2000).
\newblock Estimating synonymous and nonsynonymous substitution rates with
  realistic evolutionary models.
\newblock {\em Molecular Biology and Evolution\/}, {\bf 434}, 32--43.

\bibitem[Yang and Nielsen(2002)Yang and Nielsen]{Yang02}
Yang, Z. and Nielsen, R. (2002).
\newblock Codon-Substitution Models for Detecting Molecular Adaptation at
  Individual Sites Along Specific Lineages.
\newblock {\em Journal of Molecular Biology and Evolution\/}, {\bf 19},
  908--917.

\bibitem[Zhang {\em et~al.}(2005)Zhang, Yang, and Nielsen]{Zhang05}
Zhang, J., Yang, Z., and Nielsen, R. (2005).
\newblock Evaluation of an Improved Branch-Site Likelihood Method for Detecting
  Positive Selection at the Molecular Level.
\newblock {\em Journal of Molecular Biology and Evolution\/}, {\bf 22},
  2472--2479.

\end{thebibliography}

\end{document}